\documentclass[journal,comsoc]{IEEEtran}

\usepackage[dvips]{graphicx}
\usepackage{latexsym}
\usepackage{epsfig}
\usepackage{color}
\usepackage{amsmath} 
\usepackage{amssymb}
\usepackage{amsxtra}
\usepackage{enumitem}

\usepackage[T1]{fontenc}
\usepackage{setspace}

\usepackage{amsmath}
\usepackage[cmintegrals]{newtxmath}
\usepackage{graphicx}
\graphicspath{{Figures/}}
\usepackage{url}
\usepackage{cite}
\usepackage{algorithm}
\usepackage{algpseudocode}

\makeatletter
\def\BState{\State\hskip-\ALG@thistlm}
\makeatother

\usepackage{epstopdf}
\usepackage{amsmath}
\usepackage{amsxtra}
\usepackage{amstext}
\usepackage{amssymb}
\usepackage{latexsym}
\usepackage{dsfont} 
\usepackage{color}
\usepackage{multirow}
\usepackage{float}
\usepackage{hyperref}

\usepackage{tabularx,colortbl}
\usepackage[table]{xcolor}

\usepackage{lscape}

\usepackage{algorithm}
\usepackage{algpseudocode}

\usepackage{units}
\usepackage{cases}
\usepackage[ subrefformat=parens,labelformat=parens, caption=false, font=footnotesize]{subfig}
\usepackage{mathtools}
\mathtoolsset{showonlyrefs} 

\algnewcommand{\LineComment}[1]{\State \(\triangleright\) #1}

\newcommand{\mbf}[1]{\mathbf{#1}}

\newcommand{\nth}[1]{{#1}{\text{th}}}

\newcommand{\abs}[1]{\left|{#1}\right|}
\newcommand{\norm}[1]{\left\|{#1}\right\|}

\DeclareMathOperator*{\argmin}{arg\,min}

\usepackage{mathtools}

\newcommand{\ML}{\mathrm{ML}}

\newcommand{\NC}{\mathrm{NC}}
\newcommand{\PNC}{\mathrm{PNC}}
\newcommand{\CD}{\mathrm{CD}}
\newcommand{\PCD}{\mathrm{PCD}}

\newcommand{\SSD}{\mathrm{SSD}}

\newcommand{\SLORD}{\mathrm{SLORD}}

\newcommand{\RML}{\mathrm{RML}}
\newcommand{\RAD}{\mathrm{RAD}}

\newcommand{\err}{\mathrm{err}}

\newcommand{\F}{\mathrm{F}}

\newcommand{\Tr}{\mathrm{Tr}}
\newcommand{\Rp}{\mathring{\mbf{R}}}
\newcommand{\rp}{\mathring{r}}

\newcommand{\LoS}{\mathrm{LoS}}
\newcommand{\NLoS}{\mathrm{NLoS}}

\newcommand{\opt}{\mathrm{opt}}

\newcommand{\clu}{\mathrm{clu}}
\newcommand{\ray}{\mathrm{ray}}

\begin{document}


\title{Terahertz-Band MIMO-NOMA: Adaptive Superposition Coding and Subspace Detection}

\author{Hadi~Sarieddeen,~\IEEEmembership{Member,~IEEE,}
        Asmaa Abdallah,~\IEEEmembership{Member,~IEEE,}
        Mohammad~M.~Mansour,~\IEEEmembership{Senior Member,~IEEE}
        Mohamed-Slim~Alouini,~\IEEEmembership{Fellow,~IEEE,}
        and~Tareq~Y.~Al-Naffouri,~\IEEEmembership{Senior Member,~IEEE}

\thanks{H. Sarieddeen, M.-S. Alouini, and  T. Y. Al-Naffouri are with the Department of Computer, Electrical and Mathematical Sciences and Engineering (CEMSE), King Abdullah University of Science and Technology (KAUST), Thuwal, Makkah Province, Kingdom of Saudi Arabia, 23955-6900 (e-mail: hadi.sarieddeen@kaust.edu.sa; slim.alouini@kaust.edu.sa; tareq.alnaffouri@kaust.edu.sa).

A. Abdallah and M. M. Mansour are with the Department of Electrical
and Computer Engineering, American University of Beirut, Beirut 1107 2020,
Lebanon (e-mail: awa18@aub.edu.lb; mmansour@aub.edu.lb).

}
}

\maketitle

\begin{abstract}

We consider the problem of efficient ultra-massive multiple-input multiple-output (UM-MIMO) data detection in terahertz (THz)-band non-orthogonal multiple access (NOMA) systems. We argue that the most common THz NOMA configuration is power-domain superposition coding over quasi-optical doubly-massive MIMO channels. We propose spatial tuning techniques that modify antenna subarray arrangements to enhance channel conditions. Towards recovering the superposed data at the receiver side, we propose a family of data detectors based on low-complexity channel matrix puncturing, in which higher-order detectors are dynamically formed from lower-order component detectors. We first detail the proposed solutions for the case of superposition coding of multiple streams in point-to-point THz MIMO links. We then extend the study to multi-user NOMA, in which randomly distributed users get grouped into narrow cell sectors and are allocated different power levels depending on their proximity to the base station. We show that successive interference cancellation is carried with minimal performance and complexity costs under spatial tuning. We derive approximate bit error rate (BER) equations, and we propose an architectural design to illustrate complexity reductions. Under typical THz conditions, channel puncturing introduces more than an order of magnitude reduction in BER at high signal-to-noise ratios while reducing complexity by approximately 90\%.

\end{abstract}




\section{Introduction}


Terahertz (THz) communications \cite{Akyildiz6882305,sarieddeen2019generation} are currently accelerating research on future sixth-generation (6G) wireless mobile communications, following the successful deployment of millimeter-wave (mmWave) communications \cite{Rangan6732923} in fifth-generation (5G) networks. THz-band communications, between the microwave and optical bands, above $\unit[100]{GHz}$ and below $\unit[10]{THz}$, respectively, are celebrated as crucial enablers for ubiquitous wireless communications \cite{akyildiz2014terahertz,rajatheva2020scoring}. Recent technological advances \cite{sengupta2018terahertz} closing the THz-gap are photonic (higher data rates), electronic (higher output power), and plasmonic (especially graphene-based) \cite{6708549Jornet}. However, the research community should address several challenges before realizing the full capabilities of THz communications.

THz signal propagation incurs very high losses that severely undermine the promised gains, such as achieving a terabit-per-second (Tbps) data rate \cite{rajatheva2020scoring}. Infrastructure and algorithmic enablers should thus complement THz communications. At the infrastructure level, ultra-massive multiple input multiple output (UM-MIMO) antenna arrangements \cite{akyildiz2016realizing} and intelligent reflecting surfaces (IRSs) \cite{faisal2019ultra} are required to overcome the very short communication distances. At the algorithmic level, THz-specific signal processing techniques \cite{sarieddeen2020overview} can get around the limitation of THz quasi-optical propagation and realize seamless connectivity. In particular, optimized distance-adaptive resource allocation solutions \cite{Han7321055} are required to tackle spectrum shrinking due to molecular absorption. Furthermore, THz single-carrier modulations can replace orthogonal frequency-division multiplexing (OFDM) to reduce the baseband complexity and avoid the peak-to-average power ratio problem. Most importantly, efficient THz baseband signal processing is crucial to reduce the gap between the huge promised bandwidths and the limited state-of-the-art digital sampling speeds and processing capabilities \cite{Weithoffer8109974}.

Achieving higher spectrum utilization and enhancing the already large available bandwidths in the THz-band can be realized using advanced multiple access methods. Orthogonal multiple access (OMA) systems in which wireless resources are allocated for various users orthogonally in time, frequency, or code domains have traditionally dominated wireless communication standards. OMA's main drawback is its low spectral efficiency when allocating resources to users having poor channel conditions. Non-orthogonal multiple access (NOMA) \cite{Ding7973146,Dai8357810} is introduced to solve this problem by enabling users with significantly different channel conditions to share resources. NOMA can achieve such sharing through superposition coding (SC) of data streams at the transmitter, followed by successive interference cancellation (SIC) at the receiver. Proper user pairing criteria and power allocation policies can guarantee user fairness, affordable complexity overheads, and low signaling costs. 

While spectrum is not the main bottleneck at THz frequencies, more spectral efficiency is always desirable, especially in low-complexity single-carrier systems. Maintaining fairness among users in a congested THz band is another motive for THz NOMA, where the gap between the best and worst user scenarios is significantly large. NOMA schemes can further mitigate the hardware constraints that limit the beamforming capabilities in THz devices. Since high-frequency NOMA \cite{Zhu8798636} is likely to be conducted over line-of-sight (LoS) UM-MIMO links, we can apprehend the concept of THz multiple access through studying superposition coding of different data streams over a single point-to-point link. Very few THz-NOMA works exist in the literature. In \cite{Ulgen9298080}, the prospects of enhancing the achievable data rates in THz NOMA are highlighted. Furthermore, a bandwidth-aware THz NOMA solution is proposed in \cite{Zhang8824971}, whereas THz MIMO-NOMA energy efficiency is addressed in \cite{zhang2020energy} by optimizing power allocation, user clustering, and hybrid precoding. 


THz UM-MIMO systems will most probably follow adaptive hybrid arrays-of-subarrays (AoSA) architectures \cite{sarieddeen2020overview,Lin7786122} where each subarray (SA) typically supports independent analog beamforming. A variety of probabilistic shaping and index modulation techniques \cite{Sarieddeen8765243} can be explored for adaptive array usage, especially in plasmonic THz solutions where antenna elements (AEs) can tune the frequency of operation by simple material doping or electrostatic bias \cite{6708549Jornet}. Such reconfigurability can simultaneously accomplish resource allocation, user clustering, and beamforming. MIMO-NOMA configurations can be single-cluster or multi-cluster. In a single-cluster setting, all users except one conduct SIC, whereas in a multi-cluster setting, users are first partitioned into clusters to reduce interference, and SIC follows. The full potential of THz MIMO-NOMA is realized by the joint optimization of SIC, power allocation, beamforming, and user clustering \cite{Zhu7982784}. 

MIMO NOMA systems are also largely affected by the data detection scheme at the receiver side. In conventional massive MIMO, linear detectors are near-optimal due to channel hardening \cite{Ngo_2013}. However, correlated doubly-massive THz MIMO architectures require better-performing and more complex non-linear detectors \cite{Yang7244171}, which creates a true bottleneck at the baseband of THz systems \cite{sarieddeen2020overview}. Recently, a family of non-linear subset-stream MIMO detectors based on QR decomposition (QRD) has been proposed. The least-complex member of this family is the nulling-and-cancellation (NC) detector \cite{choi2006nulling}, followed by the chase detector (CD) \cite{waters2008chase} and the layered orthogonal lattice detector (LORD) \cite{Siti-1}, respectively. Furthermore, a less complex alternative family of subspace detectors \cite{8186206Sarieddeen,Mansour9298955} decomposes the channel using a punctured QRD, namely, WR decomposition (WRD). We argue that the latter provides a range of performance and complexity trade-offs suitable for THz UM-MIMO-NOMA scenarios.

This paper studies the downlink of THz UM-MIMO-NOMA systems, assuming adaptive hybrid AoSA beamforming, and focuses on data detection. We investigate two SC use cases, single-user, and multi-user (NOMA). In the single-user case, SC is employed to transmit multiple data streams over a point-to-point link. In the multi-user case, randomly distributed users in a cell are clustered and allocated different powers depending on their distance from a base station (BS), and users in each cluster are allocated the same time and frequency resources. We assume a single-cell scenario and neglect inter-cluster interference; at the receiver side, intra-cluster interference is suppressed via SIC. Since THz beams are highly directional, deriving the bounds on the achievable bit error rates (BERs) of the single-user case provides sufficient insight into the proposed NOMA data detectors' performance. The main paper contributions are:
\begin{enumerate}
  \item We study the performance of NOMA systems under THz-specific UM-MIMO channel models and applying recently-reported adaptive spatial tuning techniques \cite{sarieddeen2020overview} that enhance THz channel conditions.
   \item We present a family of QRD-based and WRD-based detectors tailored for THz UM-MIMO-NOMA. These detectors are extensions to reference subspace detectors that were previously proposed \cite{8186206Sarieddeen} in the context of conventional large MIMO systems. We aim to significantly lessen the baseband computational complexity and simplify THz NOMA detectors' implementation while minimizing performance loss.
  \item We propose a simple low-complexity power allocation (through AE allocation per SA) and user clustering NOMA solution that exploits the distance-based THz path-loss between the users and the BS to mitigate intra-cluster interference. 
  \item We analyze the proposed detectors' BER performance by deriving approximate closed-form equations. Without loss of generality, we analyze the proposed detectors' performance in the single-user case and empirically illustrate their scalability in NOMA settings.
  \item We propose an efficient architectural design that realizes the proposed detectors, and we study the corresponding computational complexity under THz channel conditions and Tbps baseband constraints.
\end{enumerate}

\begin{figure*}[t]
  \centering
  \subfloat[Vectorized AoSA-based single-user THz MIMO SC.]{\label{fig:model_su} \includegraphics[width=0.48\linewidth]{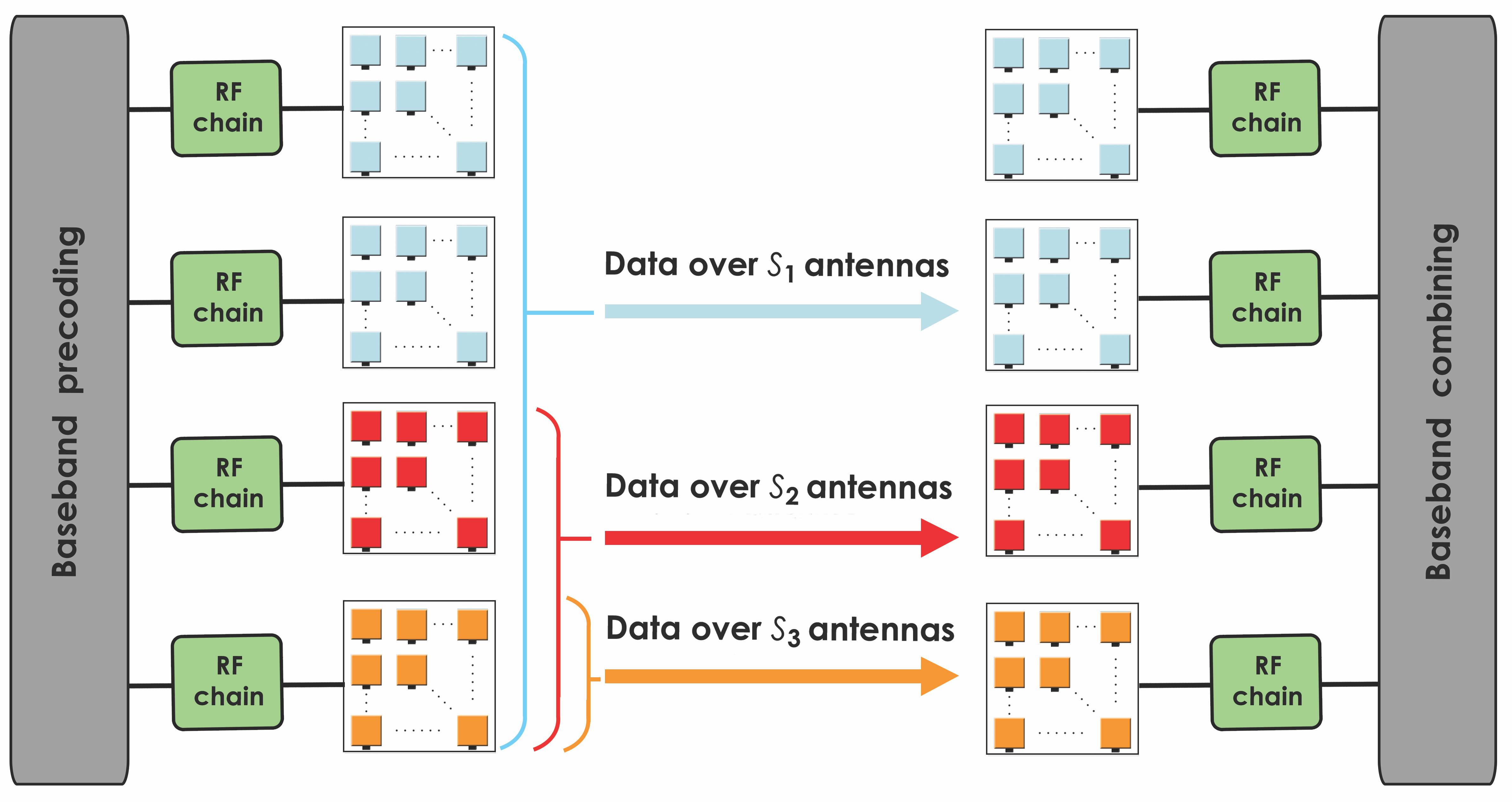}}
  \hfill
  \subfloat[Multi-user THz MIMO NOMA.]{\label{fig:model_mu} \includegraphics[width=0.43\linewidth]{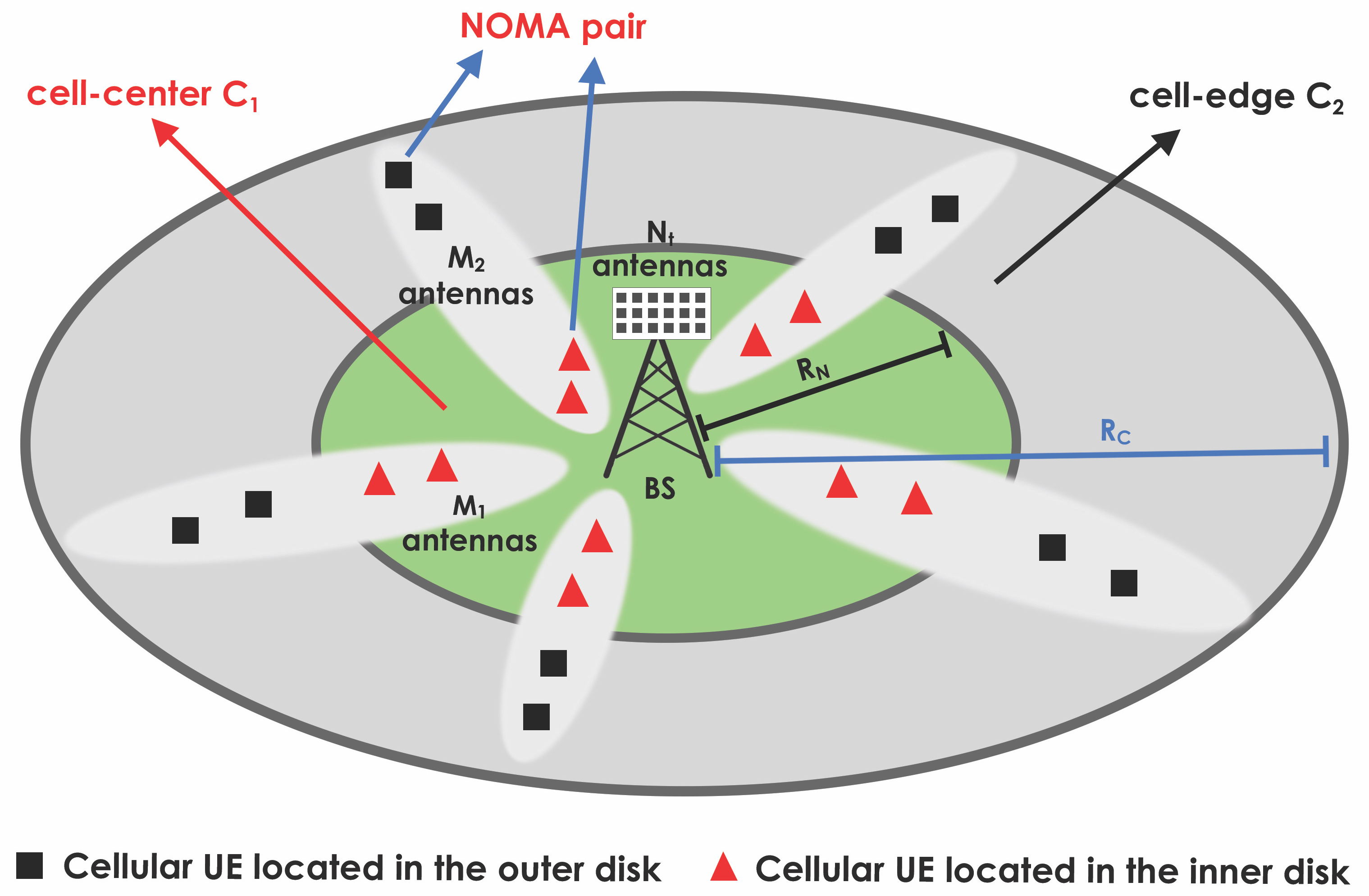}}

  \caption{System models.}
  \label{fig:sys_model}
\end{figure*}

This paper's remainder is organized as follows: We first detail the system models in Sec.~\ref{sec:sysmodel}. Then, we present THz spatial tuning techniques in Sec. \ref{sec:sp_tuning}. Afterward, we illustrate the proposed single-user detectors in Sec.~\ref{sec:single_user}, followed by the proposed NOMA clustering, power allocation, and detection schemes in Sec.~\ref{sec:multi_user}. We derive the BER equations corresponding to SIC error propagation in the proposed detectors in Sec.~\ref{sec:probBER}, and conduct the complexity study in Sec.~\ref{sec:complexity}, showcasing an efficient architecture that realizes the proposed solutions. We present the simulation results in Sec.~\ref{sec:simulations} and draw conclusions in Sec.~\ref{sec:conclusions}. Concerning notation, lower case, bold lower case, and bold upper case letters correspond to scalars, vectors, and matrices, respectively. We denote scalar norms, vector $\text{L}_2$ norms, and Frobenius norms by $\abs{\cdot}$, $\norm{\cdot}$, and $\norm{\cdot}_{\F}$, respectively. We also denot by $(\cdot)^{T}$, $(\cdot)^{*}$, $\Tr(\cdot)$, $\Re(\cdot)$, and $\mathsf{E}[\cdot]$, the transpose, conjugate transpose, trace function, real part, and expected value, respectively. $\mathcal{CN}(\cdot)$ denotes the complex normal distribution, $Q(\cdot)$ refers to the Q-function, where $Q(x)\!=\!\int_x^{\infty} e^{-z^2/2}/\sqrt{2\pi}\ dz$, and $\mathsf{P}[\cdot]$ is the probability function. $\Rp\!=\![\rp_{ij}]$ with entries $\rp_{ij}$ is a punctured matrix, and $\mbf{I}_N$ is an identity matrix of size $N$. 


\section{System Model}
\label{sec:sysmodel}


We utilize the three-dimensional (3D) THz UM-MIMO model of \cite{Sarieddeen8765243}. The AoSAs consist of $M_t\!\times\!N_t$ SAs at the transmitter and $M_r\!\times\!N_r$ SAs at the receiver. Each SA is further formed of $Q\times Q$ AEs. We denote by $\delta$ and $\Delta$ the separation between two AEs and two SAs, respectively. THz propagation is highly directional because of low reflection losses, negligible scattered and refracted components, high-gain directional antennas, and large array beamforming gains. These factors result in LoS dominance with typical survival of a single path; a ``pencil beam'' generated by each SA through analog beamforming. Therefore, we adopt a system model of LoS transmission over single-carrier frequency-flat THz channels in this paper's main embodiments. We then account for a more accurate channel model with persisting multipath components in simulation results. For simplicity, we virtually vectorize the AoSAs in the remainder of this work by setting $M_t\!\times\!N_t\!=\!N$ and $M_r\!\times\!N_r\!=\!M$. With each SA being allocated a dedicated RF chain, SAs become the smallest addressable elements of the MIMO multiplexing system, and the role of baseband precoding and combining reduces to simply defining the utilization of the SAs. 


\subsection{Use Case 1: Single-User THz MIMO SC}
\label{sec:usecase1}

In power-domain SC, we consider concurrently sending multiple data streams across various overlapping combinations of transmitting and receiving SAs (overlying channel matrices). We denote by $\mathcal{S}$ the set of superposition-coded data streams of dimensions $S_i$, $i\!=\!{1,\cdots,\abs{\mathcal{S}}}$, where $S_i\!\geq \!S_{i+1}$ and $S_1\!=\!N$. The multiplexed transmitted symbol vector $\mbf{x}_i\!=\![x^i_{1}\cdots x^i_{n}\cdots x^i_{S_i}]\!\in\!\mathcal{X}^{S_i\times1}$ is mapped to a contiguous set of antennas of indices $N\!-\!S_i\!+\!1,\cdots,N$, where $\mathcal{X}$ is a quadrature amplitude modulation (QAM) constellation. We thus have the effective channel matrices $\mbf{H}_{i}\!\in\!\mathbb{C}^{M\times S_i}$ being comprised of the columns $N\!-\!S_i\!+\!1,N\!-\!S_i\!+\!2,\cdots,N$ of the overall channel matrix $\mbf{H}\!=\![\mbf{h}_{1}\mbf{h}_{2}\cdots \mbf{h}_{N^{}}]\!\in\!\mathbb{C}^{M\times N}$. Note that we assume the SC of data symbols to happen at the higher SA indices for convenience. However, SC can still occur on arbitrary SA subsets and channel column permutations would approximate the desired structure. Furthermore, in the proposed channel-punctured solutions, the selected SAs need not be contiguous as long as one SA is common across all streams (more on that in Sec. \ref{sec:complexity}). The equivalent input-output baseband system model can then be expressed as
\begin{equation}\label{eq:sys_model_su}
\mbf{y}= \sum_{i=1}^{\abs{\mathcal{S}}} \mbf{H}_{i}\mbf{x}_i + \mbf{n},
\end{equation}
where $\mbf{y}\!=\![y_1\cdots y_{m}\cdots y_{M}]\!\in\!\mathcal{C}^{M\times1}$ is the received symbol vector and $\mbf{n}\!\in\!\mathcal{C}^{M\times1}$ is the $\mathcal{CN}(0,\sigma^2)$ noise \big($\mathsf{E}[\mbf{n}\mbf{n}^{*}]\!=\!\sigma^{2}\mbf{I}_{M}$\big). SC designates different power levels to the superposed transmitted symbol vectors. We consider allocating a higher power level $p_i$ to smaller-dimension symbol vectors $i$, i.e., $p_i\!<\!p_{i+1}$. Hence, each symbol $x^i_{n}$ is an element of a scaled complex constellation $\mathcal{X}^i$ ($\mathsf{E}[x_{n}^{i*}x_{n}^i]\!=\!p_i$), and we thus have $\mbf{x}_i\!\in\!\tilde{\mathcal{X}}^i$, where $\tilde{\mathcal{X}}^i$ is the lattice formed from all possible symbol vectors that can be generated from the $S_i$ $\mathcal{X}^i$ constellations.



The system model with $\abs{\mathcal{S}}\!=\!3$ multiplexed data streams and $N\!=\!M\!=\!8$ SAs is illustrated in Fig.~\ref{fig:model_su}, where $\mbf{x}_1$, $\mbf{x}_2$, and $\mbf{x}_3$ are transmitted from $S_1\!=\!8$, $S_2\!=\!4$, and $S_3\!=\!2$ antennas, respectively. Since the SIC order is the same as that of power levels, $\mbf{x}_3$ will be decoded first, followed by $\mbf{x}_2$ after canceling the effect of $\mbf{x}_3$, and finally $\mbf{x}_1$ after canceling the effects of $\mbf{x}_2$ and $\mbf{x}_3$.

An element of $\mbf{H}$, $h_{m,n}$, the frequency response between the $\nth{n}$ transmitting and $\nth{m}$ receiving SAs, is defined as 
\begin{equation}\label{eq:channel}
	h_{m,n} = \mbf{a}^{*}_r(\phi_r,\theta_r)G_r \alpha_{m,n} G_t \mbf{a}_t(\phi_t,\theta_t),
\end{equation} 
where $\alpha$ is the path gain, $\mbf{a}_r$ and $\mbf{a}_t$ are the receive and transmit SA steering vectors, $G_r$ and $G_t$ are the receive and transmit antenna gains, and $\phi_r$/$\theta_r$ and $\phi_t$/$\theta_t$ are the receive and transmit azimuth/elevation angles of arrival and departure, respectively. The LoS path gain is defined as
\begin{equation}\label{eq:LoS}
	\alpha_{m,n}^{\LoS} =  \frac{c}{4\pi f d_{m,n}} e^{ -\frac{1}{2} \mathcal{K}(f) d_{m,n} }  e^{  -j \frac{2\pi f}{c} d_{m,n}},
\end{equation} 
where $d_{m,n}$ is the communication distance, $c$ is the speed of light, $f$ is the center carrier frequency, and $\mathcal{K}(f)$ is the molecular absorption coefficient. $\mathcal{K}(f)$ is computed \cite{Jornet5995306} as a summation of absorption contributions from isotopes of gases in a medium. Note that we neglect the effect of mutual coupling, assuming sufficient antenna separations. The ideal analog steering vector per SA at the transmitter side is 
\begin{equation}\label{eq:steering}
	\mbf{a}_t(\phi_t,\theta_t) \!=\! \frac{1}{Q} [\!e^{j\Phi_{1,1}}\!,\!\cdots\!,e^{j\Phi_{1,Q}},e^{j\Phi_{2,1}},\!\cdots\!,e^{j\Phi_{p,q}},\!\cdots\!,e^{j\Phi_{Q,Q}}\!]\!^T,
\end{equation} 
where $\Phi_{p,q}$ is the phase shift that corresponds to AE $(p,q)$, and is defined as
\begin{align}\label{eq:shifts}
	\Phi_{p,q} &= \psi_x^{(p,q)}\frac{2\pi}{\lambda}\cos \phi_t \sin \theta_t \\&+ \psi_y^{(p,q)}\frac{2\pi}{\lambda}\sin \phi_t \sin \theta_t  + \psi_z^{(p,q)}\frac{2\pi}{\lambda}\cos \theta_t,
\end{align} 
for a wavelength $\lambda$, where $\psi_x^{(p,q)}$, $\psi_y^{(p,q)}$, and $\psi_z^{(p,q)}$ are the AE 3D coordinates. Note that we adopted a plane wave assumption for steering vectors because separations between antenna elements can be very small in plasmonic solutions \cite{6708549Jornet}.





The THz channel can still be frequency-selective, especially in indoor sub-THz scenarios where sufficient multipath components persist, although much sparser than at mmWave frequencies. The non-LoS (NLoS) component of THz multipath channels can be expressed using the Saleh-Valenzuela (S-V) model as \cite{Lin7036065}
\begin{align}
     h^{\NLoS}_{m,n} =& \sum_{v=0}^{N_{\clu}-1} \sum_{u=0}^{N_{\ray}^{(v)}} \mathbf{a}_{r}^{*}\left(\phi^{(u,v)}_{r}, \theta^{(u,v)}_{r}\right) G_{r}\left(\phi^{(u,v)}_{r}, \theta^{(u,v)}_{r}\right) \\ 
    & \alpha^{\NLoS (u,v)}_{m,n} G_{t}\left(\phi^{(u,v)}_{t}, \theta^{(u,v)}_{t}\right) 
     \mathbf{a}_{t}\left(\phi^{(u,v)}_{t}, \theta^{(u,v)}_{t}\right),
\end{align}
where $N_{\clu}$ is the number of multipath clusters and $N_{\ray}^{(v)}$ is the number of paths in the $\nth{v}$ cluster, with each path having random angles of departure and arrival within a beam region. We further have
\begin{equation}
\mathsf{E}\left[\left|\alpha^{\NLoS (u,v)}_{m,n}\right|^{2}\right]= \left(\frac{c}{4 \pi f d_{m,n}}\right)^{2}\ e^{-\mathcal{K}(f) d_{m,n}} e^{-\frac{\tau_{v}}{\Gamma}} e^{-\frac{\bar{\tau}_{v,u}}{\gamma}},
\end{equation}
where $\tau_{v}$ and $\bar{\tau}_{v,u}$ are the times of arrival (following paraboloid or exponential distributions) and ${\Gamma}$ and ${\gamma}$ are the decay factors of the clusters and rays, respectively. The angles of departure and arrival are calculated as
\begin{equation}\begin{array}{ll}
\phi^{(u,v)}_{t}=\Phi_{t}^{(u)}+\varphi_{t}^{(u,v)}, & \phi^{(u,v)}_{r}=\Phi_{r}^{(u)}+\varphi_{r}^{(u,v)} \\
\theta^{(u,v)}_{t}=\Theta_{t}^{(u)}+\vartheta_{t}^{(u,v)}, & \theta^{(u,v)}_{r}=\Theta_{r}^{(u)}+\vartheta_{r}^{(u,v)},
\end{array}\end{equation}
where $\Phi_{t}^{(u)}$/$\Phi_{r}^{(u)}$ and $\Theta_{t}^{(u)}$/$\Theta_{r}^{(u)}$ are the cluster azimuth and elevation angles of departure/arrival that follow uniform distributions over $(-\pi,\pi]$ and $\left[-\frac{\pi}{2},\frac{\pi}{2}\right]$, respectively, and  $\varphi_{t}^{(u,v)}/\varphi_{r}^{(u,v)}$ and $\vartheta_{t}^{(u,v)}/\vartheta_{r}^{(u,v)}$ are the ray azimuth and elevation angles of departure/arrival that follow a zero-mean second order Gaussian mixture model.

%



\subsection{Use Case 2: Multi-User THz MIMO-NOMA}
\label{sec:usecase2}

For NOMA, we assume the users in a cell to be divided into two groups: Users in the first group are distributed over an inner disk (${C}_1$) of radius $ R_{\mathrm{N}}$ centered at the BS, whereas users in the second group are uniformly distributed over an outer disk (${C}_2$) from $ R_{\mathrm{N}}$ to $ R_{\mathrm{C}}$. We assume a BS with $N$ SAs to service two users simultaneously and over the same frequency (power-domain SC): An $M_{1}$-SA user 1 in ${C}_1$ and $M_{2}$-SA user 2 in ${C}_2$. The received symbol vectors $\mbf{y}_1$ (at user 1) and $\mbf{y}_2$ (at user 2) are thus expressed as
\begin{align}\label{eq:sys_model}
\mbf{y}_1 &= \mbf{H}_{1}\mbf{x}_1 + \mbf{H}_{1}\mbf{x}_2 + \mbf{n}_1 \\
\mbf{y}_2 &= \mbf{H}_{2}\mbf{x}_1 + \mbf{H}_{2}\mbf{x}_2 + \mbf{n}_2,
\end{align}
where $\mbf{x}_1$ and $\mbf{x}_2$ are the transmitted power-multiplexed symbol vectors (from all $N$ SAs) and $\mbf{H}_1$ and $\mbf{H}_{2}$ are the equivalent channel sub-matrices, with $\sigma_{\mbf{H}_{1}}$ and $\sigma_{\mbf{H}_{2}}$ denoting the distance-dependent large scale fading coefficients (mainly due to path loss). Therefore, NOMA is realized by clustering inner disk users with outer disk users and designating different power levels to the transmitted superposed symbol vectors.

The single-cell multi-user MIMO-NOMA scenario is illustrated in Fig. \subref*{fig:model_mu}. We consider the number of users to be distributed according to a homogeneous Poisson point process (PPP): $\bar{\Phi}_1$ with density $\bar{\lambda}_1$ in ${C}_1$ and $\bar{\Phi}_2$ with density $\bar{\lambda}_2$ in ${C}_2$, where $\mathsf{P}[\bar{\Phi}_j\!=\!\eta] \!=\! e^{-\bar{\lambda}_j}\tfrac{\bar{\lambda}_j^{\eta}}{\eta!}$. We denote by $K$ the equal number of users in both disks, which is a Poisson random variable with mean $\mathsf{E}[K]=\bar{\lambda}_1\pi R_{\mathrm{N}}^2=\bar{\lambda}_2\pi R_{\mathrm{C}}^2-\bar{\lambda}_2\pi R_{\mathrm{N}}^2$. The total number of users is thus $2\times\mathsf{E}[K]$. Since future THz communication systems are expected to support a massive number of users/devices, we assume high user concentration. Such dense scenarios facilitate grouping users over narrow sectors as dictated by the narrow THz beamwidths.

\section{Spatial Tuning in the THz Band}
\label{sec:sp_tuning}

\begin{figure}[t]
  \centering
  \includegraphics[width=3.5in]{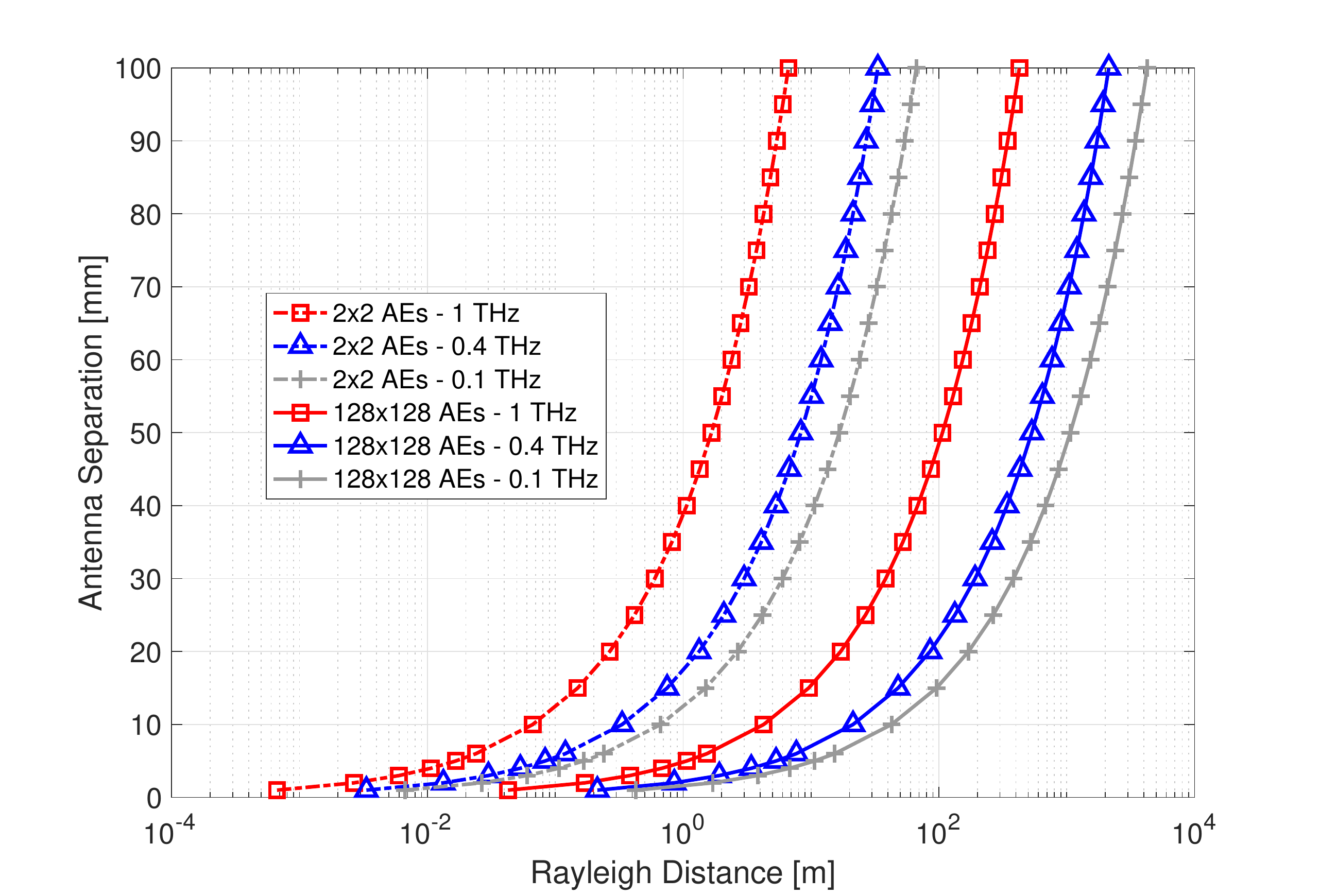}
  \caption{Rayleigh distances as a function of antenna separations for various THz LoS configurations.}\label{f:Rayleigh}
\end{figure}

UM-MIMO systems at THz frequencies are mainly employed to overcome the high absorption and propagation losses. The corresponding high channel correlation in such systems, however, limits the achievable spatial multiplexing gains. While the severity of channel correlation at THz frequencies is challenging, novel reconfigurable THz devices enable unique opportunities to deal with such correlation. In particular, by tuning the separation between SAs and AEs at the transmitter and the receiver, and without complex precoding and combining schemes, good channel conditions can be maintained \cite{Torkildson6042312}. In particular, for each communication distance ($D$), there is an optimal inter-antenna separation ($\Delta$) for which the channel is orthogonal, thus supporting a maximum number of eigenchannels over which we can transmit multiple data streams. Such dynamic tuning of antenna separations can be achieved in real-time, especially in plasmonic solutions \cite{sarieddeen2020overview}. A shorter wavelength $\lambda$ and a smaller $D$ both result in shorter optimal SA separation $\Delta_\opt$, where for symmetric MIMO systems ($M_t\!=\!N_t\!=\!M_r\!=\!N_r\!=\!M$), we have \cite{Sarieddeen8765243}
\begin{equation}
\Delta_\opt = \sqrt{zD\lambda/M},
\end{equation}
for odd values of $z$. Nonetheless, spatial tuning fails if $D$ is very large, larger than the so-called ``Rayleigh'' distance (a function of physical array dimensions) \cite{6800118Wang}. Fig. \ref{f:Rayleigh} illustrates the Rayleigh distance as a function of $\Delta$ for different frequencies and array sizes ($M\!=\!2$ and $M\!=\!128$). For very small $\Delta$s (few millimeters), a large $M$ is required to achieve a few meters of efficient communications under spatial multiplexing. Note that for the same $\Delta$, higher frequencies and more antennas extend the multiplexing-achieving distance. However, for a fixed footprint, a larger $M$ results in a Rayleigh distance reduction that is quadratic in $\Delta$. 

A massive number of AEs can be suited in a few millimeters for antenna arrays operating in the THz band. Such compactness is further emphasized with plasmonic antennas, in which $\delta$ can be reduced below $\lambda/2$ without exciting mutual coupling effects \cite{6708549Jornet}. THz spatial tuning consists of tuning $\Delta$ and calibrating the required number of AEs per SA. Tuning $\Delta$ can be achieved by maintaining a specific number of idle AEs between active SAs. In order to make better use of the idle AEs, multicarrier \cite{faisal2019ultra} configurations have been introduced (plasmonic AEs can be tuned to different frequencies without changing their physical dimensions). Furthermore, for a given communication distance, the required $Q$ AEs per SA are allocated to achieve the target beamforming gain; the possible combinations of SAs (one RF chain per SA) dictates the achievable diversity gain. We propose leveraging such adaptability for power allocation and user clustering in SC and NOMA systems, where SA dimensions (number of AEs per SA) can realize power allocation schemes. Note that in this discussion, we did not account for non-uniform array architectures, wideband channels effects, and more accurate spherical wave assumptions.






%
\section{Proposed SC Detectors}\label{sec:single_user}

This section considers the single-user SC scenario of (Sec.~\ref{eq:sys_model_su}). We propose MIMO detectors that build on three QRD-based detectors (NC, CD, and LORD) and three WRD-based detectors \cite{8186206Sarieddeen} (punctured NC (PNC), punctured CD (PCD), and the subspace detector (SSD)). While the formulations per data stream are simple extensions to \cite{8186206Sarieddeen}, under SC, we further account for proper stream ordering, sub-matrix selection, and inter-stream interference cancellation.


\subsection{Proposed QRD-Based Detectors}\label{sec:QRD}

With perfect knowledge of the channel at the receiver side, QRD decomposes $\mbf{H}$ into $\mbf{H}\!=\!\mbf{Q}\mbf{R}$, where $\mbf{Q}\!\in\!\mathcal{C}^{M \times N}$ is formed of orthonormal columns ($\mbf{Q}^*\mbf{Q}\!=\!\mbf{I}_N$), and $\mbf{R}\!=\![r_{uv}^{}]\!\in\!\mathcal{C}^{N\times N}$ is a square upper-triangular matrix (UTM) having real and positive diagonal entries. The modified baseband model is expressed as
\begin{equation}\label{eq:sysmodel2}
  \mbf{\tilde{y}} = \mbf{Q}^{*}\mbf{y} = \mbf{R}\mbf{x} + \mbf{Q}^{*}\mbf{n},
\end{equation}
with $\mbf{n}$ and $\mbf{Q}^{*}\mbf{n}$ being statistically identical.

By construction, we assume that the streams allocated higher power levels and consequently detected first at the receiver are transmitted via smaller sets of contiguous antennas, including the last SA $N$. Consequently, a single channel decomposition is sufficient to detect all streams (more on that in Sec.~\ref{sec:complexity}). After channel matrix decomposition, $\mbf{x}_{\abs{\mathcal{S}}}$ is first detected, followed by $\mbf{x}_{\abs{\mathcal{S}}-1}$, and so forth. In what follows, subscript $i$ indicates that the detection routine corresponds to detecting symbol vector $\mbf{x}_i$ of the $\nth{i}$ data stream. 

We assume optimality in the log-max sense; the maximum likelihood (ML) detector exhaustively searches the lattice $\mathcal{X}^i$ for
\begin{equation}\label{eq:ML_dist}
  \hat{\mbf{x}}_i^{\ML} = \min_{\mbf{x}_i \in \mathcal{X}^i}\norm{\mbf{y} - \mbf{H}_i\mbf{x}_i}^{2} \approx \min_{\mbf{x}_i \in \mathcal{X}^i}\norm{\mbf{\tilde{y}}_i - \mbf{R}_i\mbf{x}_i}^{2},
\end{equation}
where $\mbf{R}_i$ is the bottom right square submatrix of $\mbf{R}$ of size $S_i$, $\mbf{H}_i$ consists of the last $S_i$ columns of $\mbf{H}$, and $\mbf{\tilde{y}}_i$ consists of the last $S_i$ elements of $\mbf{\tilde{y}}$. Although this approximation does not take full advantage of receive diversity $M$, it is a key observation that allows for a cost-efficient modular architecture (Sec.~\ref{sec:complexity}). All proposed detectors will thus exploit the alternative system model 
\begin{equation}\label{eq:sysmode_alt}
  \mbf{\tilde{y}}_i = \mbf{R}_i\mbf{x}_i + \mbf{n}_i,
\end{equation}
where $\mbf{n}_i$ consists of the last $S_i$ elements of $\mbf{n}$. Note that we do not include inter-stream interference in this equation; we account for such interference in the BER analysis of Sec. \ref{sec:probBER}.

A low-complexity NC detector first performs nulling by multiplying $\mbf{y}$ with $\mbf{Q}^{*}$, which is an operation that is common to all streams, to suppress interference at layer $n$ from $x_l$ ($l >n$).  Co-antenna interference is then suppressed via back-substitution and slicing. $\hat{\mbf{x}}^{\NC}_i=[ \hat{x}^{\NC}_{1,i}\cdots\hat{x}^{\NC}_{n,i}\cdots\hat{x}^{\NC}_{S_i,i} ]$ is thus computed as
\begin{equation}\label{eq:SIC2}
\hat{x}^{\NC}_{n,i}=\left\lfloor \left(\tilde{y}_{n,i} - \sum_{l=n+1}^{S_i} r_{nl,i}\hat{x}^{\NC}_{l,i}\right)/r_{nn,i} \right\rceil_{\mathcal{X}_i},
\end{equation}
for $n\! =\!S_i,S_i\!-\!1,\cdots,1$, where $\lfloor \alpha \rceil_{\mathcal{X}} \triangleq \argmin_{x \in \mathcal{X}} \abs{\alpha-x}$ denotes the slicing operation on $\mathcal{X}$. 

With CD, the error propagation in back-substitution and slicing is mitigated by searching a reduced candidate symbol vector list $\mathcal{L}^i(\mbf{\tilde{y}}_i,\mbf{R}_i)$ before making a final decision. We first partition $\mbf{\tilde{y}}_i$, $\mbf{R}_i$, and $\mbf{x}_i$ as
\begin{equation}\label{eq:partitionCD}
    \mbf{\tilde{y}}_i =
        \begin{bmatrix}
            \mbf{\tilde{y}}_{1,i} \\
            \tilde{y}_{S_i,i}^{}
        \end{bmatrix}, \ \
    \mbf{R}_i =
        \begin{bmatrix}
            \mbf{A}_i & \mbf{b}_i \\
            \mbf{0}_i & c_i
        \end{bmatrix}, \ \
    \mbf{x}_i =
        \begin{bmatrix}
            \mbf{x}_{1,i} \\
            x_{S_i,i}^{}
        \end{bmatrix},
\end{equation}
where $\mbf{\tilde{y}}_{1,i} \!\in\! \mathcal{C}^{(S_i-1)\times1}$, $\mbf{A}_i \!\in\! \mathcal{C}^{(S_i-1)\times(S_i-1)}$, $\mbf{b}_i \!\in\! \mathcal{C}^{(S_i-1)\times1}$, $c_i \!\in\! \mathcal{R}^{1\times1}$, $\mbf{x}_{1,i} \!\in\! {\mathcal{X}^i}^{S_i-1}$, and $\mbf{0}_i$ is a $1\!\times\!(S_i\!-\!1)$ zero-valued vector. For each root-layer $x_{S_i,i}$ value, a candidate vector is constructed as in~\eqref{eq:SIC2} and appended to $\mathcal{L}^i$. After populating $\abs{\mathcal{X}^i}$ candidate vectors, the final hard-output (HO) solution is selected from $\mathcal{L}^i$ as
\begin{equation}\label{eq:sol_CD}
  \hat{\mbf{x}}^{\CD}_i = \argmin_{\mbf{x}_i \in \mathcal{L}^i} \norm{\mbf{\tilde{y}}_i\!-\!\mbf{R}_i\mbf{x}_i}^{2}.
\end{equation}

By repeating the CD routine, LORD iterates chase detection over various layer orderings, for different root layers, by shifting the columns of $\mbf{H}$ cyclically and accumulating the root-layer symbol of every CD output. Every permuted $\mbf{H}$ at step $t$, $t\!=\!1,\!\cdots\!,S_i$, is QR-decomposed into $\mbf{Q}^{(t)}$ and $\mbf{R}^{(t)}$ following~\eqref{eq:partitionCD}. Denote by $\hat{\mbf{x}}_{i,(t)}^{\CD}$ the CD output at step $t$. The overall LORD solution is
\begin{align}\label{eq:LORD}
\hat{\mbf{x}}_i^{\SLORD}&=[\hat{x}^{\SLORD}_{1,i}\! \cdots\! \hat{x}^{\SLORD}_{n,i}\! \cdots\hat{x}^{\SLORD}_{S_i,i}] \\ \hat{x}^{\SLORD}_{n,i} &= \hat{x}_{S_i\!-\!n\!+\!1,i(t=n)}^{\CD}.
\end{align}

\subsection{Proposed WRD-Based Detectors}\label{sec:WRD}

Channel puncturing can significantly reduce the complexity of QRD-based detectors. WRD transforms $\mbf{H}$ into a punctured UTM $\Rp=[\rp_{uv}]\in\mathcal{C}^{N\times N}$ with $\rp_{uu}\in\mathcal{R}^{+}$ by zeroing-out the entries between column $N$ and the diagonal, via a matrix multiplication $\mbf{W}^{*}\mbf{H}=\Rp$, where $\mbf{W}\in\mathcal{C}^{M\times N}$. The brute-force procedure for computing $\mbf{W}$ \cite{ojard2008method} requires complex matrix inversions that are also prone to roundoff errors. Nevertheless, a simpler alternative procedure \cite{2014_mansour_SPL_WLD} applies QRD followed by elementary matrix operations. The modified symbol vector at the receiver is 
\begin{equation}\label{eq:sysmodel3}
  \mbf{\bar{y}} = \mbf{W}^{*}\mbf{y} = \Rp\mbf{x} + \mbf{W}^{*}\mbf{n}.
\end{equation}
We similarly define $\Rp_i$ as the bottom right square submatrix of size $S_i$ of $\mbf{R}$, and $\mbf{\bar{y}}_i$ as the last $S_i$ elements of $\mbf{\tilde{y}}$. Then, by analogy with \eqref{eq:partitionCD} we have for the $\nth{i}$ stream
\begin{equation}\label{eq:partitionCDwrd}
\mbf{\bar{y}}_i =
    \left[\begin{IEEEeqnarraybox*}[][c]{c}
        \mbf{\bar{y}}_{1,i} \\
        \bar{y}_{S_i,i}^{}
    \end{IEEEeqnarraybox*}\right], \ \
\Rp_i =
    \begin{bmatrix}
        \mathring{\mbf{A}}_i & \mathring{\mbf{b}}_i \\
        \mbf{0}_i & \mathring{c}_i
        \end{bmatrix}, \ \  \mbf{x}_i=\begin{bmatrix} \mbf{x}_{1,i} \\ x_{S_i,i}^{}  \end{bmatrix},
\end{equation}
where in this case $\mathring{\mbf{A}}_i \in \mathcal{R}^{(S_i-1)\times(S_i-1)}$ is diagonal. 

With PNC, we pre-multiply with $\mbf{W}^{*}$ instead of $\mbf{Q}^{*}$ for nulling, and perform back-substitution and slicing as
\begin{equation}\label{eq:SICpunctured}
    \hat{x}^{\PNC}_{n,i} = \left\lfloor \left(\bar{y}_{n,i} - \rp_{nS_i,i}^{}\hat{x}_{S_i,i}^{\PNC}\right)/\rp_{nn,i} \right\rceil_{\mathcal{X}^i},
\end{equation}
for $n\! =\!S_i,S_i\!-\!1,\cdots,1$, where $\hat{\mbf{x}}^{\PNC}_i\!=\![ \hat{x}^{\PNC}_{1,i}\!\cdots\!\hat{x}^{\PNC}_{n,i}\cdots\hat{x}^{\PNC}_{S_i,i} ]$ and $\hat{x}^{\PNC}_{S_i,i} \!=\! \left\lfloor \bar{y}_{S_i,i}/\rp_{S_iS_i,i}\right\rceil_{\mathcal{X}^i}$. For all streams, slicing on layers $n \!=\! S_i\!-\!1,\cdots, 1$ is executed in parallel because $\mathring{\mbf{A}}_i$ is diagonal.

PCD performs the chase detection operations following the partition in \eqref{eq:partitionCDwrd}. An altered list of relevant symbol vectors $\mathcal{P}^i(\mbf{\bar{y}}_i,\Rp_i)$ is thus populated, and the corresponding distance to a vector $\mbf{x}_i=[\mbf{x}_{1,i},x_{S_i,i}]^T$ is given by
\begin{equation}\label{eq:xwr}
    \bar{d}(\mbf{x}_i)    \!=\!
        \norm{\mbf{\bar{y}}_i\!-\!\Rp_i\mbf{x}_i}^{2} \!=\! \abs{\bar{y}_{S_i,i}^{}\!-\!\mathring{c}_i x_{S_i,i}^{}}^{2}\!+\!\norm{\mbf{\bar{y}}_{1,i}\!-\!\mathring{\mbf{A}}_i\mbf{{x}}_{1,i}\!-\!\mathring{\mbf{b}}_ix_{S_i,i}^{}}^{2}.
\end{equation}
For every $x_{S_i,i}^{}\in\mathcal{X}_i$, this distance is minimized as
\begin{align}\label{eq:xwr2}
    \min_{\mbf{x}_{1,i}\!\in\! \mathcal{X}^{S_i\!-\!1}} \! \bar{d}(\mbf{x}_i)
        \!&=\! \abs{\bar{y}_{S_i,i}^{}\!-\!\mathring{c}_i x_{S_i,i}^{}}^{2}\!+\!
              \min_{\mbf{x}_{1,i}\!\in\! \mathcal{X}^{S_i\!-\!1}}\! \norm{\mbf{\bar{y}}_{1,i}\!-\!\mathring{\mbf{A}}_i\mbf{{x}}_{1,i}\!-\!\mathring{\mbf{b}}_ix_{S_i,i}^{}}^{2}\\
        &= \abs{\bar{y}_{S_i,i}^{}\!-\!\mathring{c}_i x_{S_i,i}^{}}^{2} \!+\!
              \norm{\mbf{\bar{y}}_{1,i}\!-\!\mathring{\mbf{A}}_i\mbf{\hat{x}}_{1,i}(x_{S_i,i})\!-\!\mathring{\mbf{b}}_ix_{S_i,i}^{}}^{2} \\
              &\triangleq \bar{d}^*\left(\mbf{x}_i(x_{S_i,i})\right),
\end{align}
where $\mbf{\hat{x}}_{1}(x_{S_i,i})=\lfloor(\mbf{\bar{y}}_{1,i}\!-\!\mathring{\mbf{b}}_i x_{S_i,i}^{})/\mathring{\mbf{A}}_i\rceil_{\mathcal{X}^{S_i-1}}$, which is a vectorized slicing operation, and $\mbf{x}_i(x_{S_i,i})=[\mbf{\hat{x}}_{1,i}(x_{S_i,i}),x_{S_i,i}]^T$. We then add the symbol vector $\mbf{x}_i(x_{S_i,i})$ to $\mathcal{P}^i$ and save the corresponding $\bar{d}^*\left(\mbf{x}_i(x_{S_i,i})\right)$. The HO solution $\hat{\mbf{x}}_i^{\PCD}$ is selected from $\mathcal{P}^i$ as the vector with the minimum distance.

SSD picks from the PCD HO vector the symbol at the root layer for each step $t$. Therefore, the SSD HO symbol vector is assembled over $N$ executions of PCD, one symbol at a time, as
\begin{align}\label{eq:SSDout}
\hat{\mbf{x}}_i^{\SSD} &=[\hat{x}^{\SSD}_{1,i}\cdots\hat{x}^{\SSD}_{n,i}\cdots\hat{x}^{\SSD}_{S_i,i}] \\ 
\hat{x}^{\SSD}_{n,i} &= \hat{x}_{S_i\!-\!n\!+\!1,i(t=n)}^{\PCD}.
\end{align}

Note that for all proposed detectors in the single-user scenario, we detect $\mbf{x}_{\abs{S_i}}$ by treating the interference caused by other streams as unknown. Every time a symbol vector $\mbf{x}_{i}$ is detected, by treating streams $i-1$ down to $1$ as unknown interference, the received signal component due to $\mbf{x}_i$ gets canceled. This paves the way to detecting $\mbf{x}_{i-1}$ from the remaining part of the received signal in the next step. In the particular case of SSD, the received vector is updated before every step $i$ as follows: 
\begin{equation}\label{eq:SIC3}
\mbf{y} \leftarrow \mbf{y} - \mbf{H}_{i}\hat{\mbf{x}}_i^{\SSD}.
\end{equation}

%
\section{Extensions to Multi-User MIMO-NOMA}
\label{sec:multi_user}

Having detailed the proposed detectors, we next study their utilization in a NOMA setting. We start by proposing a low-complexity joint clustering and power control mechanism. 

\subsection{Joint Clustering and Power Control}

Although NOMA settings result in intra-cluster interference (ICI), efficient user clustering enhances ICI cancellation in SIC at the receiver. The SIC process distinguishes same-cluster users by the difference in their power, where users are allocated power levels based on their corresponding channel vector norms. Hence, an efficient clustering approach couples two users with significantly different channel vector norms, typically a user far from the BS with a near user. Motivated by this realization, we propose a low-complexity joint distance-based (path-loss-based) clustering and power control scheme (JDCP). We assume sufficiently dense networks that guarantee a sufficient number of users in a beamwidth-limited cell sector. 

The proposed clustering approach operates as follows: First, the farthest user in disk $C_1$ is grouped with the farthest user in disk $C_2$. Then, the second farthest user in $C_1$ is grouped with the second farthest user in $C_2$, and so on. Under such pairing, SIC efficiency is guaranteed because we always allocate more power to the weak user.
SIC decoding is only needed at the receiver of the strong NOMA user. User 1 with better channel conditions is the strong user, and user 2 is the weak user (the SNR at user 1 is higher than that at user 2). Hence, we have $\sigma_{\mbf{H}_{1}}^2 \geq \sigma_{\mbf{H}_{2}}^2$, which indicates that the central user is user 1 and the cell-edge user is user 2. Therefore, user 2 will be allocated more power. Subsequently, user 2 directly decodes its own data $\mbf{x}_2$, treating the interference from $\mbf{x}_1$ as unknown, while user 1 applies SIC to cancel out $\mbf{x}_2$ before decoding its own symbol vector $\mbf{x}_1$.

Following clustering, the proposed low-complexity power control (PC) mechanism exploits the cellular link's CSI to minimize the interference between NOMA pairs. We select the transmit power of NOMA pairs based on channel conditions; specifically, the distance-based path-loss. At THz frequencies, the distance-based path-loss includes the distance-based absorption loss in addition to propagation losses. However, instead of using \eqref{eq:LoS}, we consider the equivalent THz path-loss model that accounts for additional losses in the path-loss exponent. Reported LoS path-loss exponent values at sub-THz frequencies are around $\dot{\alpha} \!=\! 2.2$ \cite{Abbasi9135643}. The allocated power for the $\nth{k}$ close user (in $C_1$), based on channel inversion, is given by
\begin{equation}
p_1^{(k)}= \rho_\mathrm{rx} d_{1,k}^{\dot{\alpha}},
\end{equation}
where $d_{1,k}$ is the distance separating the BS and the $\nth{k}$ user equipment (UE) in $C_1$, $\dot{\alpha}$ is the path-loss exponent, and $\rho_\mathrm{rx}$ is the minimum required power for UE signal recovery (also referred to as receiver sensitivity). As for the power allocated to the $\nth{k}$ far user in $C_2$, it can be expressed as 
\begin{equation}
p_2^{(k)} = \min\{\mu\rho_\mathrm{rx} d_{2,k}^{\dot{\alpha}},\frac{P_{\max}}{N}-p_1^{(k)}\},
\end{equation}
where $\mu$ is a NOMA PC parameter, $d_{2,k}$ is the distance between the $\nth{k}$ UE and the BS in $C_2$, and $P_{\max}$ is the maximum transmit power.

The adopted channel inversion technique does not compensate for small-scale fading which is negligible at THz frequencies; it only accounts for the large-scale path-loss effects. Consequently, the proposed PC scheme does not require establishing instantaneous CSI at the transmitter, which is costly at high frequencies and massive dimensions. Moreover, the BS can accurately estimate distances via location updates as defined in the 3GPP TS 23.032: Universal geographical area description (GAD) ~\cite{3gpp032}. We further argue that the raging accuracy is much higher with THz signals \cite{sarieddeen2019generation}. Furthermore, this scheme is particularly suitable for SIC decoding since it guarantees allocating much more power to far users and much less power to close users, which guarantees alluding the worst-case scenario of allocating equal power for both users, a scenario that must be avoided in NOMA. The JDCP scheme is summarized in Algorithm~\ref{alg:JCPC}.

Spatial domain multiplexing of multi-clusters and multi-cells can further result in multi-cluster interference (MCI) and multi-cell interference (MCeI). However, the probability of such interference is very low at THz frequencies due to shorter communications distances and narrower beams. Coordinated beamforming techniques can be used, alongside intra-cluster SIC, to suppress intra-cluster interference, inter-cluster interference, and multi-cell interference.

\begin{algorithm}[t]
\caption{\small Joint Distance-based Clustering and Power Control}\label{alg:JCPC}\small
\begin{algorithmic}[1]
\Procedure{JDCP}{}
\LineComment{BS has distance vectors $\mbf{d}_{C_1}=[d_{1,1},... , d_{1,k},... ,d_{1,K}]$
 and $\mbf{d}_{C_2}=[d_{2,1},... , d_{2,k},... ,d_{2,K}]$  for users in $C_1$ and $C_2$, respectively.}
\LineComment{Initialize index vectors: $I_1=\{1,2,...,K\}$ for users in $C_1$ 

and $I_2=\{1,2,...,K\}$ for users in $C_2$}
\LineComment{Initialize $P_{\max}$, and $CL=\emptyset$ }
\State   LOOP: \textbf{While} {$I_1 \neq \emptyset$ and $I_2 \neq \emptyset$ do}
\State   Clustering:
\State    \:\: Choose $i_1=  \arg \max\limits _{k\in I_1 } [\mbf{d}_{C_1}]_{k}$
\State    \:\: Choose $i_2=  \arg \max\limits _{k \in I_1 } [\mbf{d}_{C_2}]_{k}$
\State    \:\: Fix $CL\gets CL \cup [i_1, i_2]$
\State  Power Allocation:
\State    \:\: $p_1^{(k)} \gets \rho_\mathrm{rx} {\{[\mbf{d}_{C_1}]_{i_1}\}}^{\dot{\alpha}}$
\State    \:\: $p_2^{(k)} \gets \min\{\mu\rho_\mathrm{rx} {\{[\mbf{d}_{C_2}]_{i_2}\}}^{\dot{\alpha}},\frac{P_{\max}}{N}-p_1^{(k)}\}$
\State   Save $I_1 \gets I_1 \setminus i_1$
\State   Save $I_2 \gets I_2 \setminus i_2$
\State       \textbf{goto} LOOP
\State \textbf{end}
\EndProcedure
\end{algorithmic}
\end{algorithm}

\subsection{Multi-User MIMO-NOMA Detection}

Following JDCP, user 2 decodes its symbol vector $\mbf{x}_2$ directly by treating the interference due to $\mbf{x}_1$ as unknown interference. Such detection can be achieved by using any of the proposed detectors in Sec.~\ref{sec:single_user}, with computations corresponding to the case of the first detected stream ($i\!=\!\abs{S}\!=\!2$). This concludes the operations at user 2 where no SIC is required. Hence, the detection routine at user 2 is, in fact, regular MIMO detection. Nevertheless, SIC-based detection applies to user 1. First, the symbol vector $\mbf{x}_2$ is detected while treating $\mbf{x}_1$ as unknown interference (since we assume no communication between users, where each user decodes its own information independently). Then, user 1 cancels the portion of the received signal that is caused by $\mbf{x}_2$ and decodes $\mbf{x}_1$ from the remainder of the received signal:

\begin{equation}\label{eq:SIC4}
\mbf{y}_1 \leftarrow \mbf{y}_1 - \mbf{H}_{1}\mbf{x}_2 = \mbf{H}_{1}\mbf{x}_1 + \mbf{n}_1.
\end{equation}
By analogy with the construction in Sec.~\ref{sec:single_user}, the operation at user 1 can be modeled as dual-stream detection ($\abs{\mathcal{S}}=2$). The difference here is that the output at the second iteration ($i=1$) is the only desired output and the output $\mbf{x}_2$ at the first iteration is discarded. 


%
\section{Characterization and Analysis of BER}
\label{sec:probBER}

Since this work's primary objective is to investigate the performance of detection schemes at the receiving side without optimizations at the transmitter side, the suitable metric for performance analyses is BER rather than achievable sum rates. In what follows, we formulate approximate BER equations that provide insight into the resulting system performance of the proposed QRD-based and WRD-based NOMA detectors. We consider single-user SC and assume the case where all streams are of equal size ($S_i\!=\!N$ for all $i$). We assume a Gaussian channel case to derive closed-form BER equations, and we generate empirical approximate BER bounds for THz channels. The relative BER performances of QRD-based and WRD-based detectors are studied in \cite{8186206Sarieddeen} for OMA-MIMO systems. The main factors that affect the performance under puncturing are: The reduction in error propagation over MIMO layers, the variation in the statistical properties of the elements of $\Rp$ compared to $\mbf{R}$, and noise colorness. In this section, we assume the generic case of detecting stream $i$ after canceling the interference from stream $i\!+\!1$. We drop the index $i$ from the symbols for clarity of presentation.


\subsection{NC and PNC}
\label{sec:berSIC}

Let $P_n^{}(r_{nn}^{})$ and $\acute{P}_n^{}(\rp_{nn}^{})$ be the bit error probabilities conditioned on $r_{nn}$ and $\rp_{nn}$, for detecting $x_n$ ($1\!\leq\!n\!\leq\!N$), with NC and PNC, respectively. For PNC, assuming scaled binary phase shift keying (BPSK), where $\mathcal{X}_i=\{-\sqrt{p_i},\sqrt{p_i}\}$, we have at layer $N$
\begin{equation}\label{eq:qfun1}
    \acute{P}_N^{}(\rp_{NN}^{}) = Q \left( \sqrt{ \frac{ 2\rp_{NN}^2p_i }{ \sigma^{2} } } \right).
\end{equation}
For other layers, if slicing at layer $N$ is accurate, we have $\acute{P}_n(\rp_{nn}) = Q \left( \sqrt{ \frac{ 2\rp_{nn}^2p_i }{ \sigma^{2} } } \right)$. Otherwise, we get
\begin{equation}\label{eq:cancellation}
\bar{y}_{n}\!-\! \rp_{nN}^{}\hat{x}_{N}^{\NC} = \rp_{nn}x_{n} + \rp_{nN}^{}(x_{N}^{} - \hat{x}_{N}^{\NC}) + w_n,
\end{equation}
where $w_n$ is the $\nth{n}$ component of the noise vector $\mbf{W}^{*}\mbf{n}$. Noting that $x_{N}^{}\!-\!\hat{x}_{N}^{\NC}\!=\!\pm 2\sqrt{p_i}$, the variance of interference plus noise is $\sigma^{2}\!+\!4p_i$. 


This analysis holds when the streams (or users in the NOMA scenario) are adequately separated in power, such that no error propagates from stream $i\!+\!1$ to $i$ and the interference from stream $i\!-\!1$ is negligible. This assumption is made by most studies on MIMO-NOMA that consider capacity maximization. However, an additional error component is introduced by inter-stream SIC. Assuming that a one-bit slicing error occurs when detecting stream $i\!+\!1$ (with Gray mapping), the corresponding error component is proportional to $\beta_{i+1} \!=\! \frac{2\sqrt{p_{i+1}}}{\log_2{L}-1}$ for a scaled $L$-QAM $\mathcal{X}_{i+1}$. Furthermore, the interference power from user $i\!-\!1$ (neglectig farther streams) is proportional to $p_{i-1}\sigma_{\mbf{H}_{i-1}}^2$. Hence, we have $\acute{P}_n(\rp_{nn}) = Q \left(  \sqrt{ \frac{ 2\rp_{nn}^2p_i }{ \sigma^{2} + p_{i-1}\sigma_{\mbf{H}_{i-1}}^2 + \acute{P}_n^{(i+1)}\sigma_{\mbf{H}_{i+1}}^2\beta_{i+1}^2 + 4p_i } } \right) $, and the resultant BER when detecting $x_n$ can be expressed as
\begin{align}\label{eq:berPSIC}
     &\acute{P}_n(\rp_{nn}) \!=\! Q \left(\! \sqrt{ \frac{ 2\rp_{NN}^2p_i }{ \sigma^{2} \!+\! p_{i-1}\!\sigma_{\mbf{H}_{i-1}}^2 \!+\! \acute{P}_n^{(i+1)}\sigma_{\mbf{H}_{i+1}}^2\!\beta_{i+1}^2 } } \!\right) \!\left(\!1\!-\!\acute{P}_N^{}(\rp_{NN}^{})\!\right) \\ &+ Q \left(  \sqrt{ \frac{ 2\rp_{nn}^2p_i }{ \sigma^{2} + p_{i-1}\sigma_{\mbf{H}_{i-1}}^2 + \acute{P}_n^{(i+1)}\sigma_{\mbf{H}_{i+1}}^2\beta_{i+1}^2 + 4p_i } } \right)  \acute{P}_N^{}(\rp_{NN}^{}).
\end{align}
The QRD-based NC BER at layer $n$ ($n\!=\!N\!-\!1,\cdots,1$) is studied in \cite{choi2006nulling}. By analogy, and following a similar derivation to that for PNC, we have 
\begin{equation}\label{eq:berSIC}
P_n(r_{nn}) = \sum_{\psi_{n+1}\in D_{n+1}} P_n(\err|r_{nn},\psi_{n+1})P_{n+1}(\psi_{n+1})
\end{equation}
\begin{align}\label{eq:berSIC2}
&P_n(\err|r_{nn},\psi_{n\!+\!1}) \!=\!  \\ &Q \!\left(\! \sqrt{\! \frac{ 2r_{nn}^2p_i }{ \sigma^{2} \!+\! p_{i\!-\!1}\sigma_{\mbf{H}_{i-1}}^2 \!+\! \acute{P}_n^{(i\!+\!1)}\!\sigma_{\mbf{H}_{i+1}}^2\beta_{i+1}^2 \!+\! 4p_i\psi_{n\!+\!1}\psi_{n\!+\!1}^{T} } } \!\right),
\end{align}
where $P_{n+1}(\psi_{n+1})$ is recursively obtained, and $\psi_n$ is an occurrence of $D_n$, the complete set of possible error patterns up to layer $n$. Since $\abs{\acute{D}_n} = 2 < \abs{D_n} = 2^{N-n+1}$ with WRD, error propagation is significantly lessened. Nevertheless, this does not guarantee an enhanced BER performance. For a Gaussian channel, puncturing reduces error propagation but also results in performance degradation. At layer $n$, the BER is derived by taking the average over $r_{nn}^2$ and $\rp_{nn}^2$. For Gaussian channels, the off-diagonal elements of $\mbf{R}$ are circular symmetric complex Gaussian random variables, and the square of the $\nth{n}$ diagonal elements is chi-squared distributed with $2(N-n+1)$ degrees of freedom. Although the distributions of non-zero off-diagonal elements remain intact in $\Rp$, the distributions of diagonal elements at upper layers $n\!=\!1,\cdots,N\!-\!3$ lose degrees of freedom from $2(N\!-\!n\!+\!1)$ down to $4$. Since $\rp_{nn}^2$ is on average smaller than $r_{nn}^2$ for $1\!\leq\!n\!\leq\!N\!-\!2$ (lower degrees of freedom), PNC results in performance loss. However, both NC and PNC are dominated by $\acute{P}_N^{}(\rp_{NN}^{})=P_N^{}(r_{NN}^{})$ at the root layer.

For a THz channel, however, we distinguish between two cases: The case of spatial tuning and the case of highly correlated channels. Under spatial tuning, the channel is nearly diagonal (up to some quantization errors), and the studied punctured, and unpunctured detectors reduce to the same detector. On the other hand, under high channel correlation, the diagonal elements of $\mbf{R}$ do not initially possess high degrees of freedom to lose them through puncturing. Hence, the reduction in error propagation across symbol layers is further emphasized in correlated THz scenarios, where puncturing could even result result in performance enhancement. Note that both configurations, with or without spatial tuning, are somewhat deterministic. Hence, no averaging is required over channel realizations; we could directly simulate empirical BER results following equations \eqref{eq:berPSIC} and \eqref{eq:berSIC}, for example. Nevertheless, in an indoor THz scenario with sufficient multipath components, the proposed detector's performance can still mimic the Gaussian case.

\begin{figure*}[t]
  \centering

  \subfloat[Gaussian $8\times8$ MIMO channel.]{\label{theo1} \includegraphics[width=0.46\linewidth]{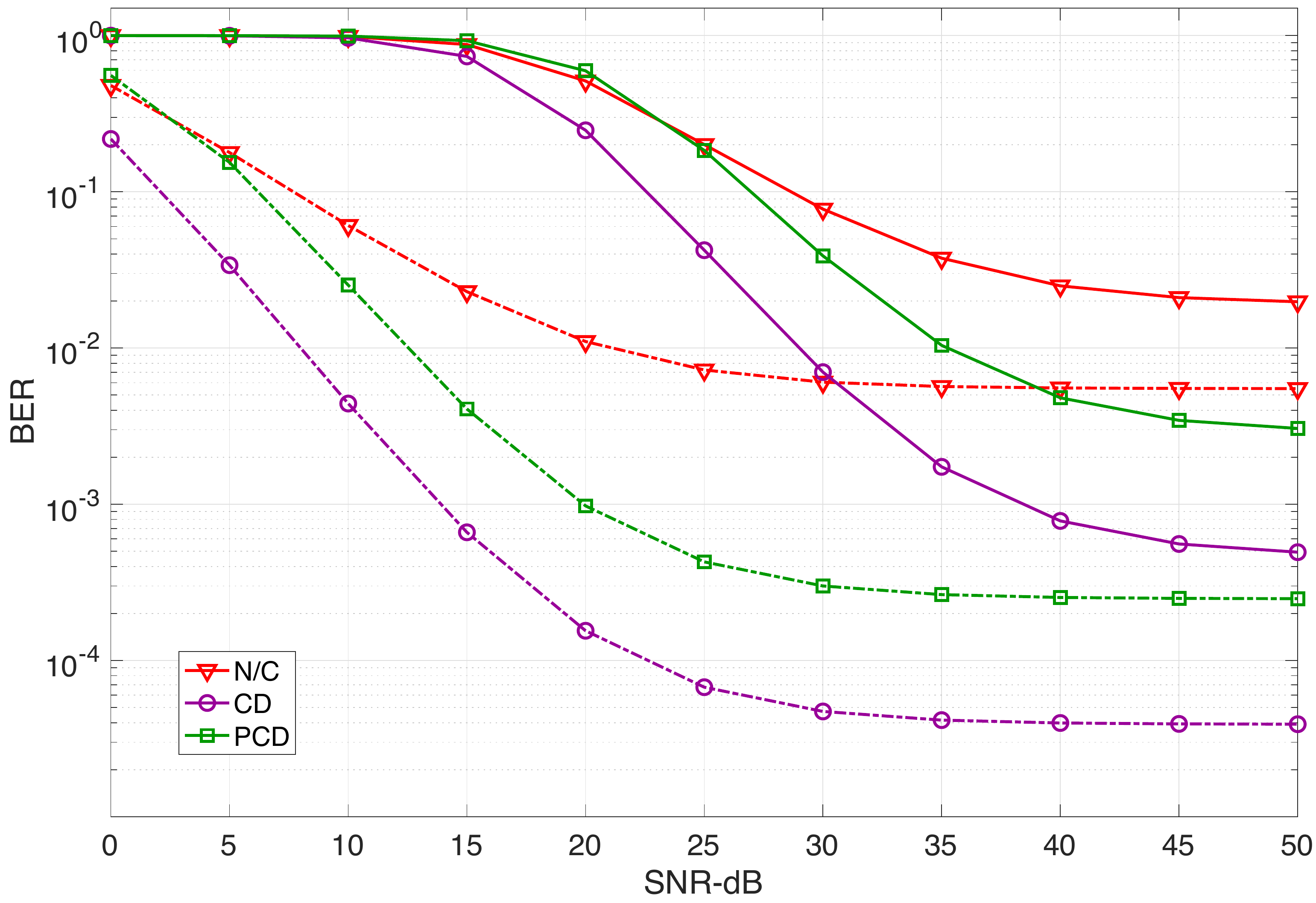}}
  \hfill
  \subfloat[Multipath THz $4\times4$ MIMO channel.]{\label{theo2} \includegraphics[width=0.52\linewidth]{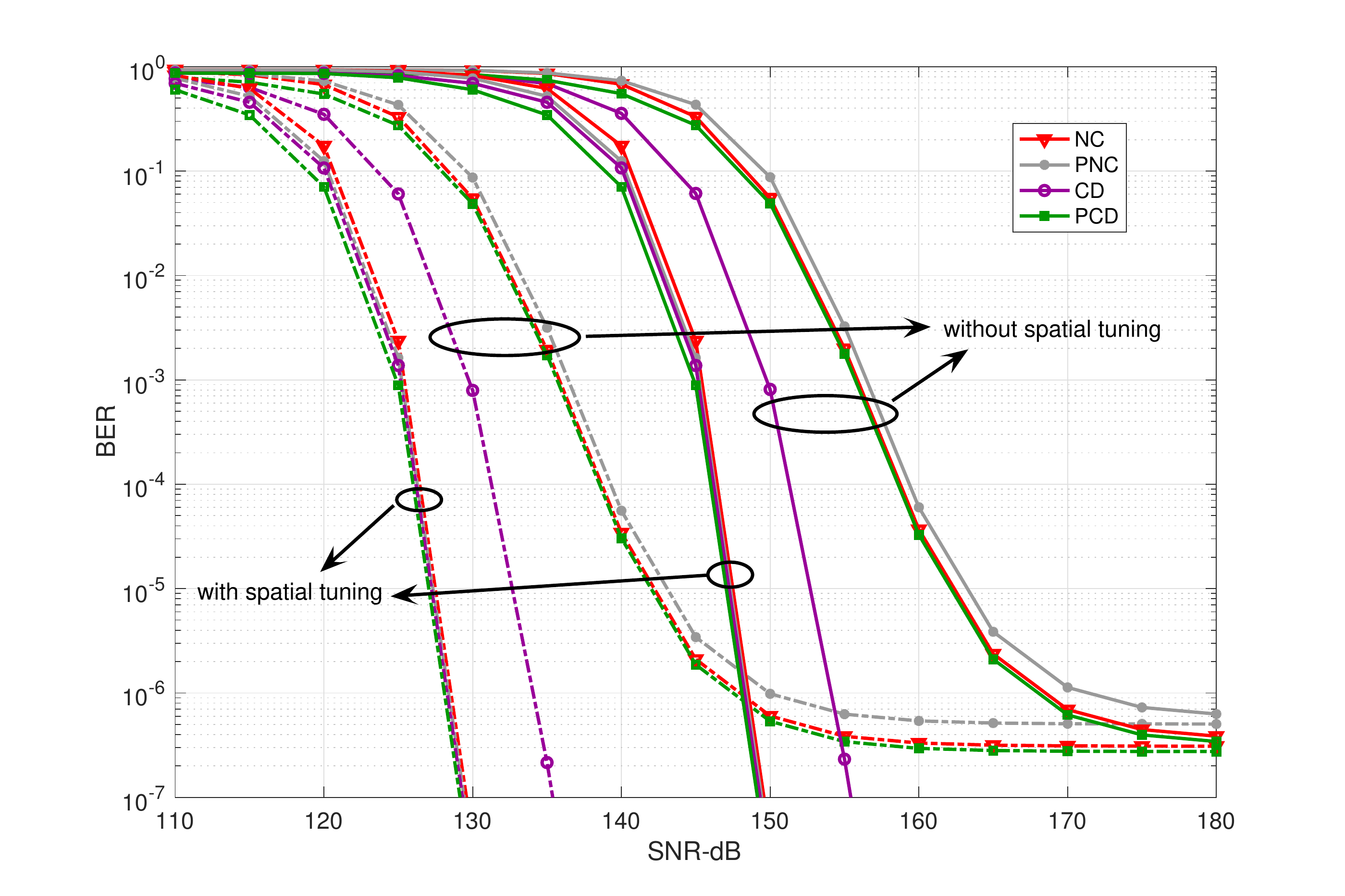}}

  \caption{Theoretical BER performance in a three-stream SC scenario (power separation in the order of 100 and assuming QPSK,  $f=1 \textrm{\ THz}$, and $D=5 \textrm{\ m}$), with solid lines denoting stream 3 and dotted lines denoting stream 2.}
  \label{sim_results_theo}
\end{figure*}

We next assume Rayleigh fading, where we can derive closed-form BER equations and extend the analysis to arbitrary modulation types. We denote by $G(z,\bar{\gamma},L)$ the function that produces the average BER for an L-QAM constellation over $z$-fold diversity Rayleigh fading with mean branch SNR $\bar{\gamma}$ \cite{Kim_2008}. The average PNC BER for layer $1\!\leq\!n\!\leq\!N\!-\!1$ is
\begin{align}\label{eq:ber_avg_1}
     \acute{P}_n &=  G \left( 2, \frac{p_i}{ \sigma^{2} \!+\! p_{i-1}\sigma_{\mbf{H}_{i-1}}^2 \!+\! \acute{P}_n^{(i+1)}\sigma_{\mbf{H}_{i+1}}^2\beta_{i+1}^2},L \right) (1-\acute{P}_N) \\&+ G \left( 2, \frac{p_i}{ \sigma^{2} \!+\! p_{i-1}\sigma_{\mbf{H}_{i-1}}^2 \!+\! \acute{P}_n^{(i+1)}\sigma_{\mbf{H}_{i+1}}^2\beta_{i+1}^2 + \beta_{i}^2},L \right) \acute{P}_N,
\end{align}
where $\beta_{i} \!=\! \frac{2\sqrt{p_{i}}}{\log_2{L}-1}$, and because of puncturing, the layers $1\!\leq\!n\!\leq\!N\!-\!1$ only provide $2$-fold diversity. Furthermore, the average NC BER at layer $n<N$ is derived by replacing $P_n(\err|r_{nn},\psi_{n+1})$ in equation \eqref{eq:berSIC} by its average over $r_{nn}$, $P_n(\err|\psi_{n+1})$, where
\begin{align}\label{eq:ber_avg_3}
    &P_n(\err|\psi_{n+1}) = \mathsf{E}[P_n(\err|r_{nn},\psi_{n+1})] \\ &=\!G\!\left(\! N\!-\!n\!+\!1,\! \frac{p_i}{ \sigma^{2} \!+\! p_{i-1\!}\sigma_{\mbf{H}_{i-1}}^2 \!+\! \acute{P}_n^{(i\!+\!1)}\!\sigma_{\mbf{H}_{i+1}}^2\beta_{i+1}^2 \!+\! \beta_{i}^2\psi_{n\!+\!1}\psi_{n\!+\!1}^{T}}\! \right).
\end{align}
Since $p_i\!+\!1$ in $\beta_{i+1}$ is much larger than $p_i$, the residual error from SIC, when it occurs, is much more severe than the residual error from back-substitution and slicing. Nevertheless, setting $p_{i+1}\! \gg \!p_i$ and $p_i\! \gg \!p_{i-1}$ renders SIC error and interference negligible, respectively. This is because $\acute{P}_n^{(i\!+\!1)}$ will approach zero in this case despite the increase in $\beta_{i+1}$, and the interference component $p_{i-1}\sigma_{\mbf{H}_{i-1}}^2$ will also be negligible (we seek maximal power separation between same-cluster users).

\subsection{CD and PCD}\label{sec:berCD}

An approximate approach for capturing the BER performance of the CD builds on the NC BER equations. Since no error propagates from layer $N$ when searching all its candidate symbols, we sum the BER combinations on layers $n<N$ assuming $\acute{P}_N^{}(\rp_{NN}^{}) = P_N^{}(r_{NN}^{}) = 0$. In the particular case of Gaussian channels, the BER of PCD is given by 
\begin{equation}\label{eq:ber_PCD}
     P^{\PCD} =  (N-1)\  G \left( 2, \frac{p_i}{ \sigma^{2} \!+\! p_{i-1}\sigma_{\mbf{H}_{i-1}}^2 \!+\! \acute{P}_n^{(i+1)}\sigma_{\mbf{H}_{i+1}}^2\beta_{i+1}^2},L \right).
\end{equation}

The approximate BER performances of NC, PNC, CD, and PCD are shown in Fig. \ref{sim_results_theo}, for a three-stream SC MIMO scenario. The reduction in complexity under puncturing comes at a graceful performance cost, and inter-stream interference results in error floors. However, under spatial tuning, all detectors behave identically and are robust to inter-stream interference. Note that without spatial tuning, channel correlation results in severe performance degradation (more results in Sec. \ref{sec:simulations}).

For LORD and SSD, a comparative BER analysis in the context of OMA-MIMO is conducted in \cite{8186206Sarieddeen}. It is argued that with correlated channels, SSD outperforms LORD at high SNR. By only considering the root-layer symbols of the PCD solution after cyclically shifting the layers of $\mbf{H}$ in each SSD iteration, intra-channel interference is mitigated (under puncturing, all layers are only dependent on the root layer). Therefore, using SSD instead of LORD not only reduces complexity but also enhances performance. Note that we considered a low complexity puncturing mechanism in this work, which does not necessarily guarantee maximum achievable rates. In \cite{Mansour9298955}, an augmented channel is punctured instead of the true channel, where the augmentation accounts for a minimum mean square error (MMSE) prefiltering and channel gain compensation. The resultant scheme maximizes the achievable rates at an additional complexity cost.





%
\section{Architecture and Complexity Analysis}\label{sec:complexity}

\begin{table*}[!t]
\centering
\footnotesize
\caption{Complexity savings in the studied detectors} \vspace{-0.1in}
\label{table:detectors} 
\centering 
\begin{tabular}{| c || c | c | c | c |} 
\hline
Detector & QRD  & Puncturing  & Savings (flops) \\
\hline\hline
NC $\rightarrow$ PNC & $\epsilon_2$ & $\epsilon_3$ & $J\times\abs{\mathcal{S}}\times \epsilon_1$ \\\hline %
CD $\rightarrow$ PCD & $\epsilon_2$ & $\epsilon_3$ & $J\times\abs{\mathcal{S}}\times \abs{\mathcal{X}_i}\times\epsilon_1$ \\\hline
LORD $\rightarrow$ SSD & $N\times\epsilon_2$ & $N\times\epsilon_3$ & $J\times\abs{\mathcal{S}}\times N\times\abs{\mathcal{X}_i}\times\epsilon_1$ \\
\hline 
\end{tabular}
\end{table*}

This section details an efficient architecture that realizes our proposed detectors in a modular and low complexity design. We similarly assume the case of detection at the $\nth{i}$ steam in a single-user SC setting with $S_i\!=\!N$, and drop the index $i$ for convenience. Figure \ref{fig:archfig} illustrates the architectural design for WRD-based detectors. The complexity reduction is on multiple levels:
\begin{enumerate}
  \item A single channel matrix decomposition (one decomposition for PNC and CD and $N$ decompositions for SSD) is required for all streams, as all subsequent streams are assumed to use contiguous subsets of transmitting SAs. Consequently, a global channel matrix QRD/WRD can be stored in hardware, and a multiplexer would select the required channel at input $i$. In the specific case of WRD, we can relax the constraint on the antenna subsets being contiguous to the condition of only including the $\nth{N}$ SA in all streams. Arbitrary combinations of SAs are thus tolerated. This observation is valid because all layers other than $N$ are independent under puncturing and can thus be flexibly arranged in any order. 
  
  \begin{figure}[t]
  \centering
  \includegraphics[width=3.2in]{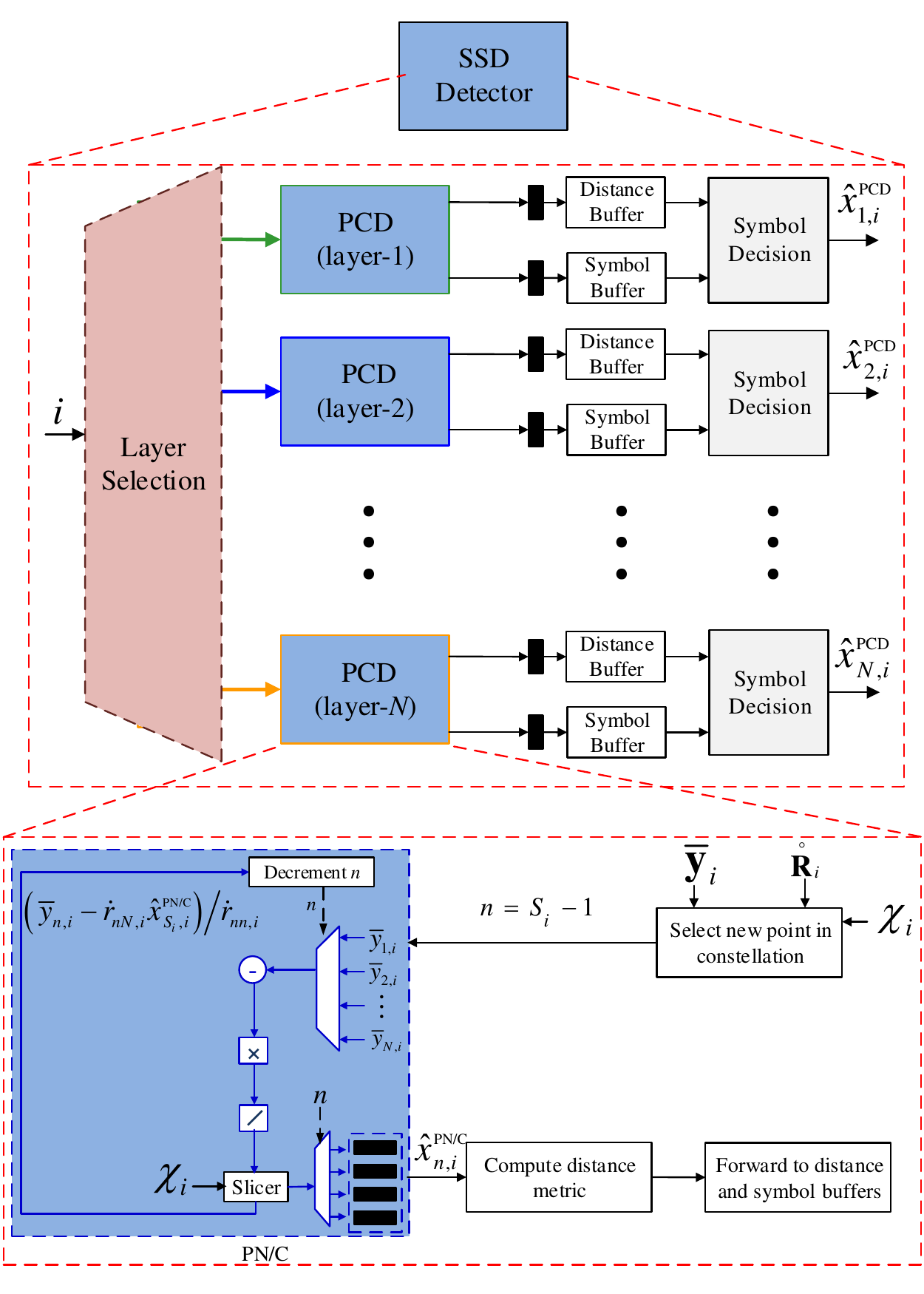}
  \caption{Hierarchical architectural design of SSD using PCDs and PNCs as building blocks.}
  \label{fig:archfig}
\end{figure}

  \item The choice of detectors, whether QRD-based or WRD-based, is such that higher complexity detectors can make use of their lower-complexity counterparts as building blocks. Hence, we propose a hierarchical architectural design in Fig.~\ref{fig:archfig}, where SSD uses PCD components that themselves use PNC. With SSD, the computed PCD distances at a layer of interest are forwarded, alongside the corresponding symbol vectors, to a decision processing unit. This occurs on all layers in parallel, where the aggregate output vector is ready after a full-layer processing delay. Similarly, a QRD-based design features LORD using CD and NC as building blocks, but the resulting architecture is not fully parallelizable. The proposed design provides the flexibility to adapt detector types depending on varying channel conditions or resource requirements while using a single dedicated hardware processor.
  \item Channel puncturing reduces the computational complexity. The PNC routine, which gets executed the most as the lowest level building block, requires reduced back-propagation computations due to puncturing-induced sparsity (significantly less complex than NC). 
  \item In the specific case when symbols on each layer are chosen from the same unscaled modulation $\mathcal{X}$ for each stream/user, we can store in memory an exhaustive set of the products $\mbf{R}\mbf{x}$ or $\Rp\mbf{x}$ for all $\mbf{x}$. Then, a simple scaling by $p_i$ at layer $i$ would replace the matrix-vector multiplication. This feature is more useful when all combinations of symbol vectors are required, which is more feasible with relatively low-order MIMO-NOMA systems. 
\end{enumerate}

We next analyze the detectors' complexity in terms of floating-point operations (flops), as a function of real addition ($\RAD$) and real multiplication ($\RML$) operations. When $\Rp\mbf{x}$ is executed in lieu of $\mbf{R}\mbf{x}$, a reduction of $(N-2)(N-1)/2$ multiplications is noted, which is equivalent to $\epsilon_1=(N^2-3N+2)\,\RAD+(2N^2-6N+4)\,\RML$ flops. This reduction accounts to $77\%$ and $88\%$ of the multiplications in a $16\!\times\!16$ MIMO system and $32\!\times\!32$ MIMO system, respectively. However, QRD consumes $\epsilon_2=(4N^3-N^2-N)\,\RAD+(4N^3+3N^2)\,\RML$ flops, and puncturing requires $\epsilon_3=\frac{2}{3}(8N^3-15N^2+4N-12)\,\RAD+(\frac{16}{3}N^3-7N^2+\frac{8}{3}N-20)\,\RML$ flops. Hence, in a NOMA setting, PNC saves $\abs{\mathcal{S}}\times \epsilon_1$ flops, PCD saves $\abs{\mathcal{S}}\times\abs{\mathcal{X}_i}\times\epsilon_1$ flops, and SSD saves $\abs{\mathcal{S}}\times N\times\abs{\mathcal{X}_i}\times\epsilon_1$ flops compared to regular MIMO QRD-based detectors. Table \ref{table:detectors} summarizes these results. With flat fading THz channels, such decomposition computations can be stored in memory for a large number of frames $J$.


\begin{figure*}[t]
  \centering
  \subfloat[$S_{2}=S_{1}=4$ - $f=1 \textrm{\ THz}$ - $D=5 \textrm{\ m}$ - Gaussian channel - QPSK - power separation in the order of 1000 (solid lines for stream 1 and dotted lines for stream 2).]{\label{fig1} \includegraphics[width=0.43\linewidth]{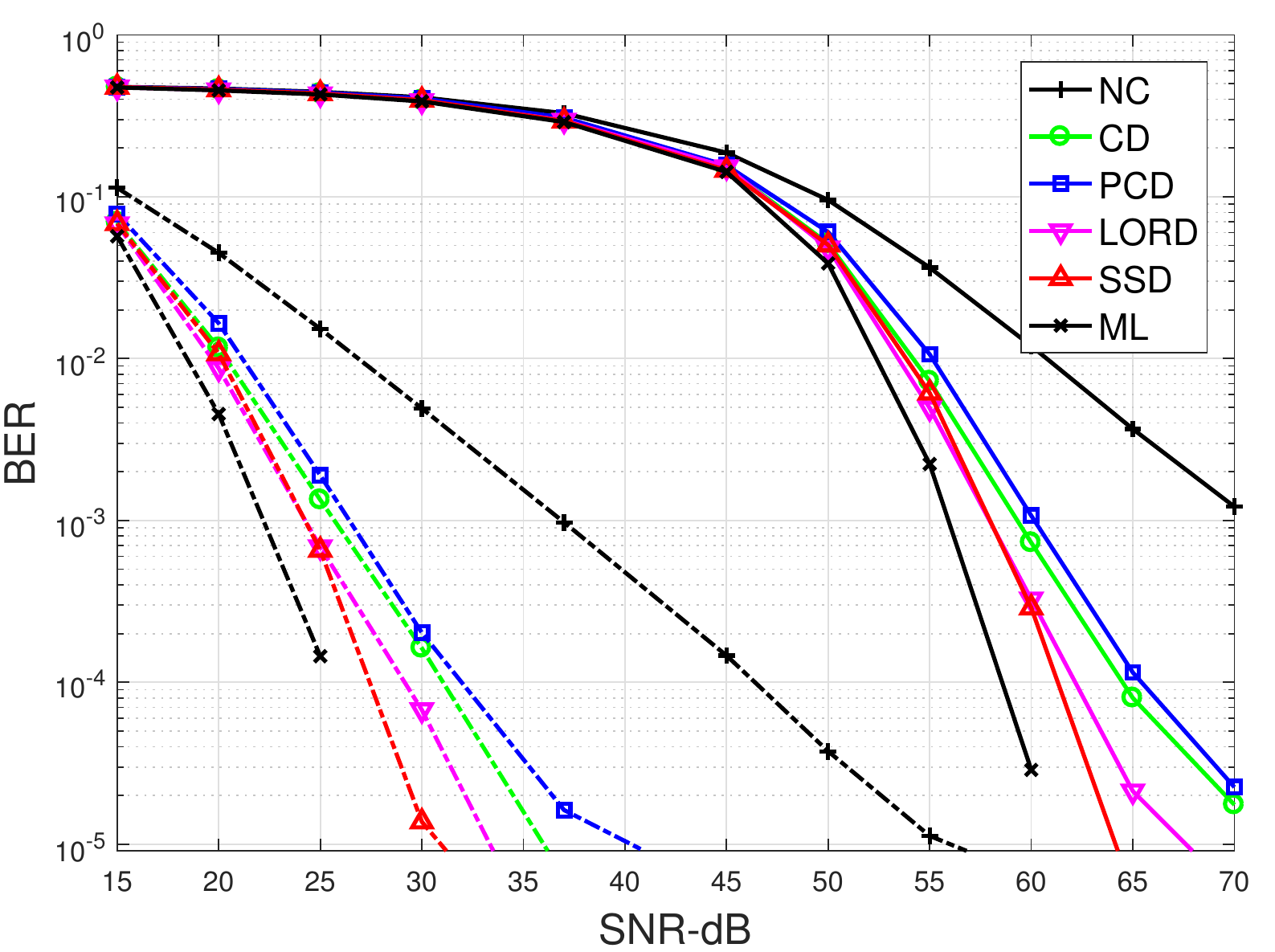}}
  \hfill
    \subfloat[$S_{3}=16$, $S_{2}=8$, $S_1=4$ - $f=0.3 \textrm{\ THz}$ - $D=5 \textrm{\ m}$ - THz multipath channel without spatial tuning - 16QAM - power separation in the order of 1000.]{\label{fig3} \includegraphics[width=0.48\linewidth]{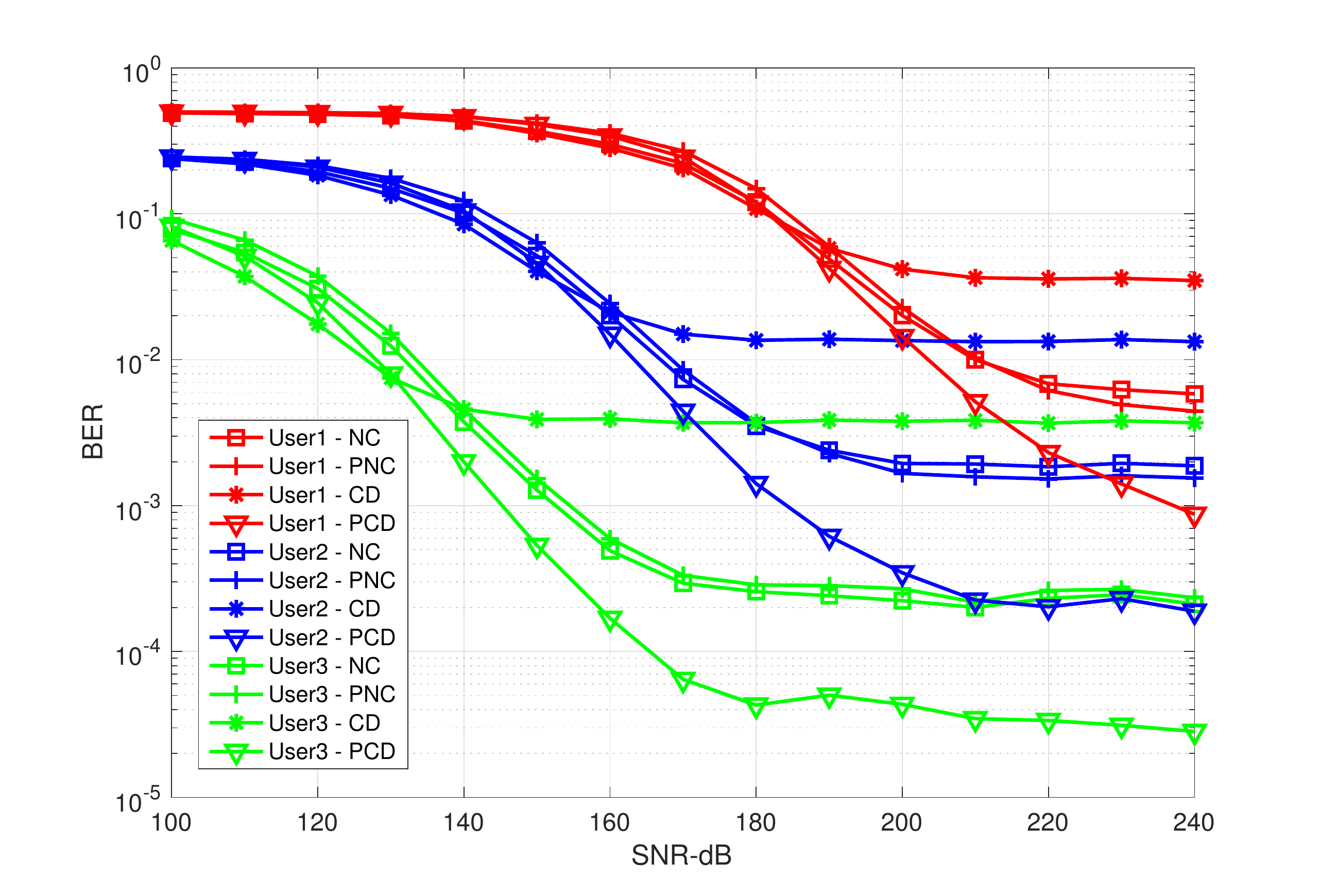}}
 
 \subfloat[$S_{3}=16$, $S_{2}=8$, $S_1=4$ - $f=1 \textrm{\ THz}$ - $D=5 \textrm{\ m}$ - LoS THz channel with spatial tuning - 16QAM - power separation in the order of 100.]{\label{fig2} \includegraphics[width=0.48\linewidth]{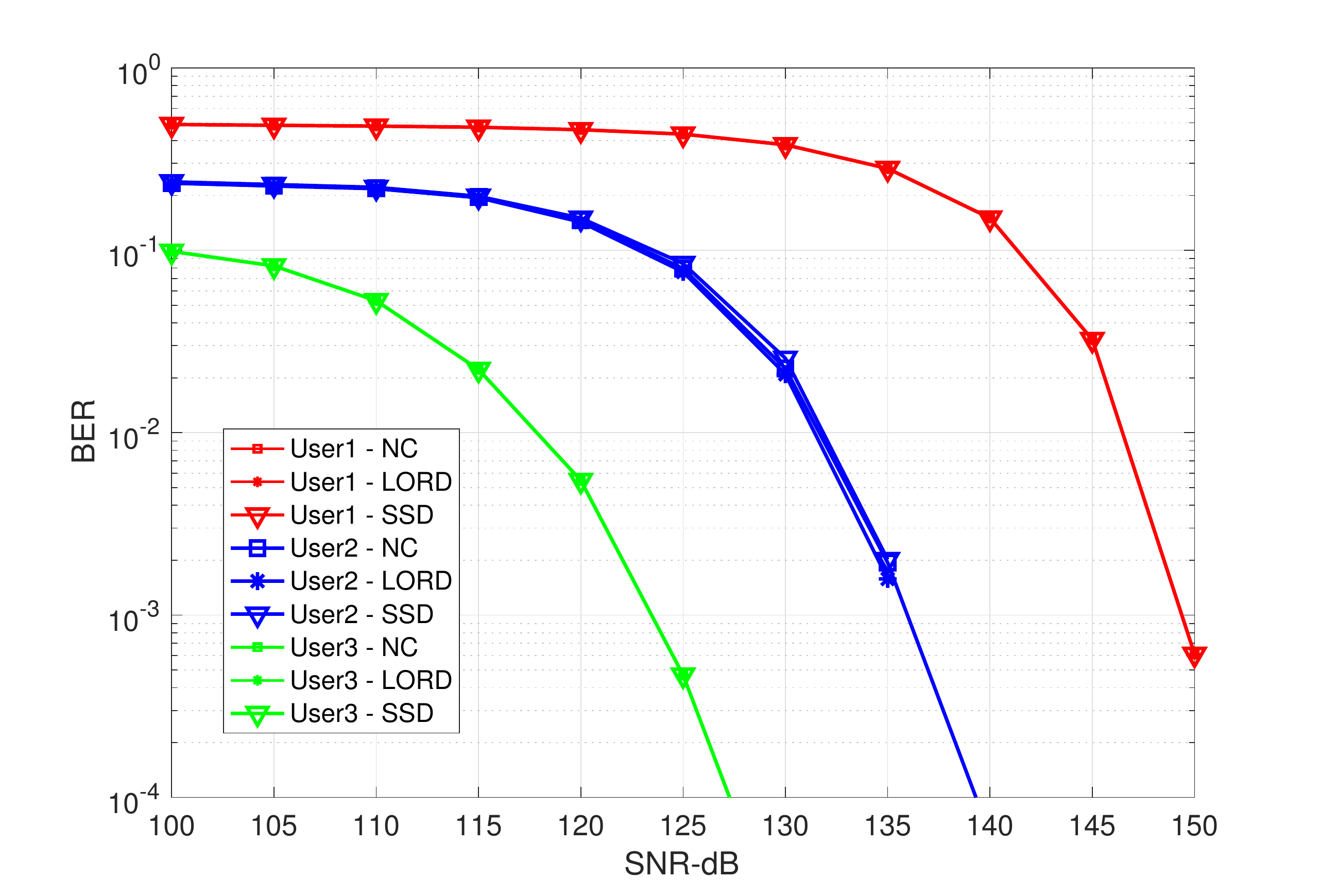}}
  \hfill
  \subfloat[$S_{3}=16$, $S_{2}=8$, $S_1=4$ - $f=1 \textrm{\ THz}$ - $D=5 \textrm{\ m}$ - LoS THz channel without spatial tuning - 16QAM - power separation in the order of 1000.]{\label{fig4} \includegraphics[width=0.48\linewidth]{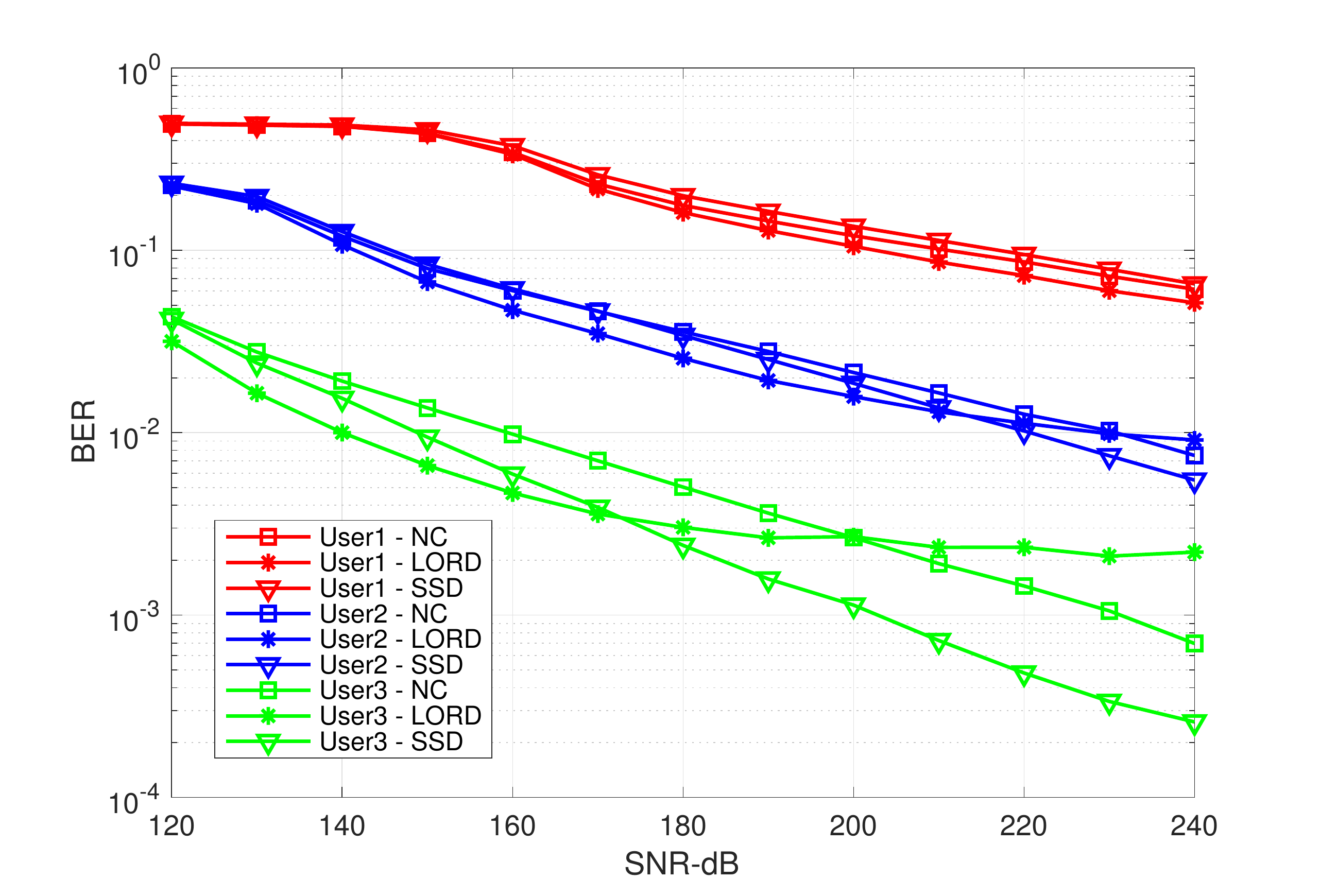}}

  \caption{BER performance of proposed detectors in the single-user scenario.}
  \label{sim_results_1}
\end{figure*}

\begin{figure*}[t]
  \centering
  \subfloat[$N=M_{1}=M_{2}=16$ - THz channel with spatial tuning for both users.]{\label{ber4} \includegraphics[width=0.48\linewidth]{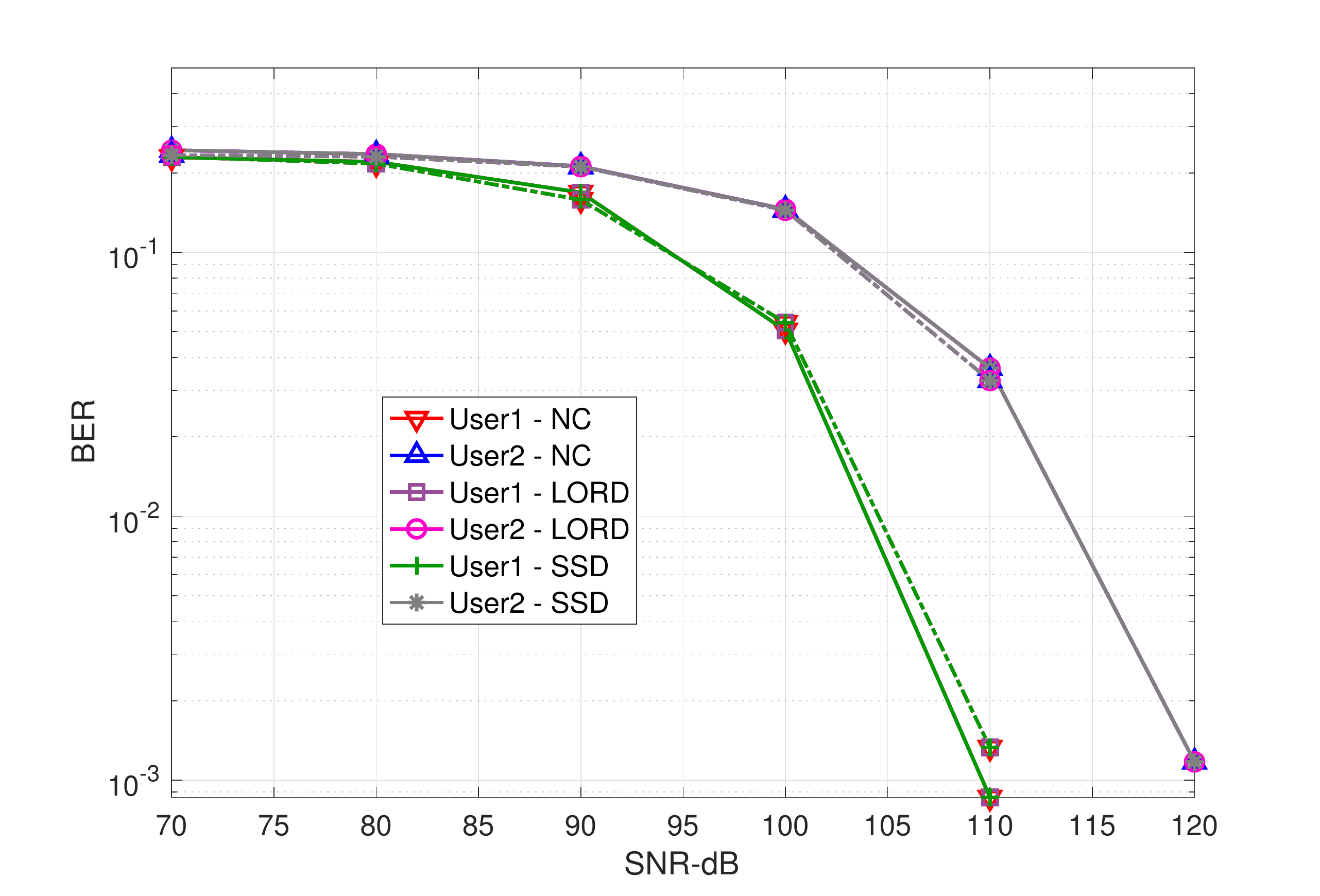}}
  \hfill
  \subfloat[$N=M_{1}=M_{2}=16$ - THz channel without spatial tuning.]{\label{ber3} \includegraphics[width=0.48\linewidth]{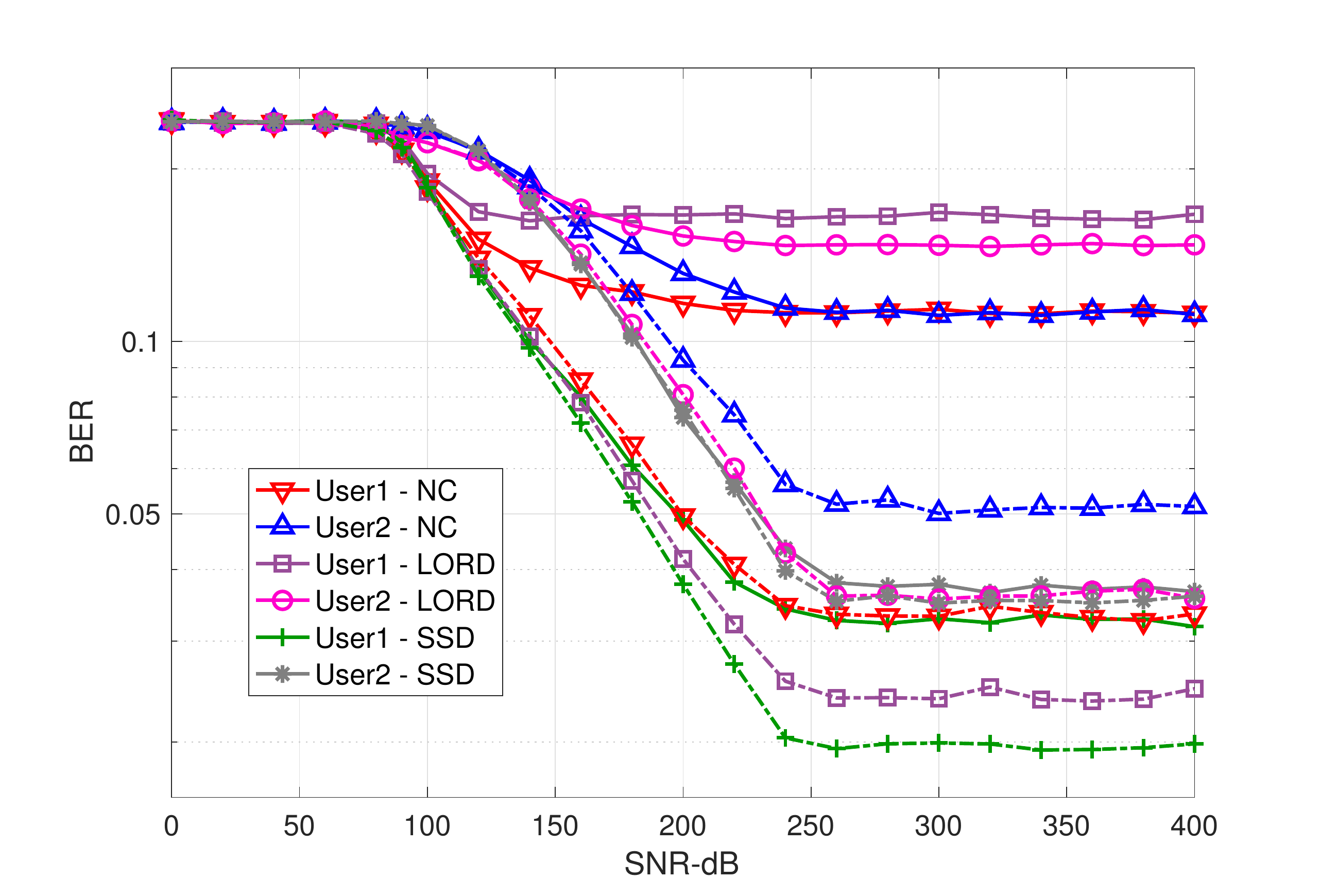}}
  
  \subfloat[$N=M_{1}=M_{2}=16$ - THz channel with spatial tuning for far user.]{\label{ber2} \includegraphics[width=0.48\linewidth]{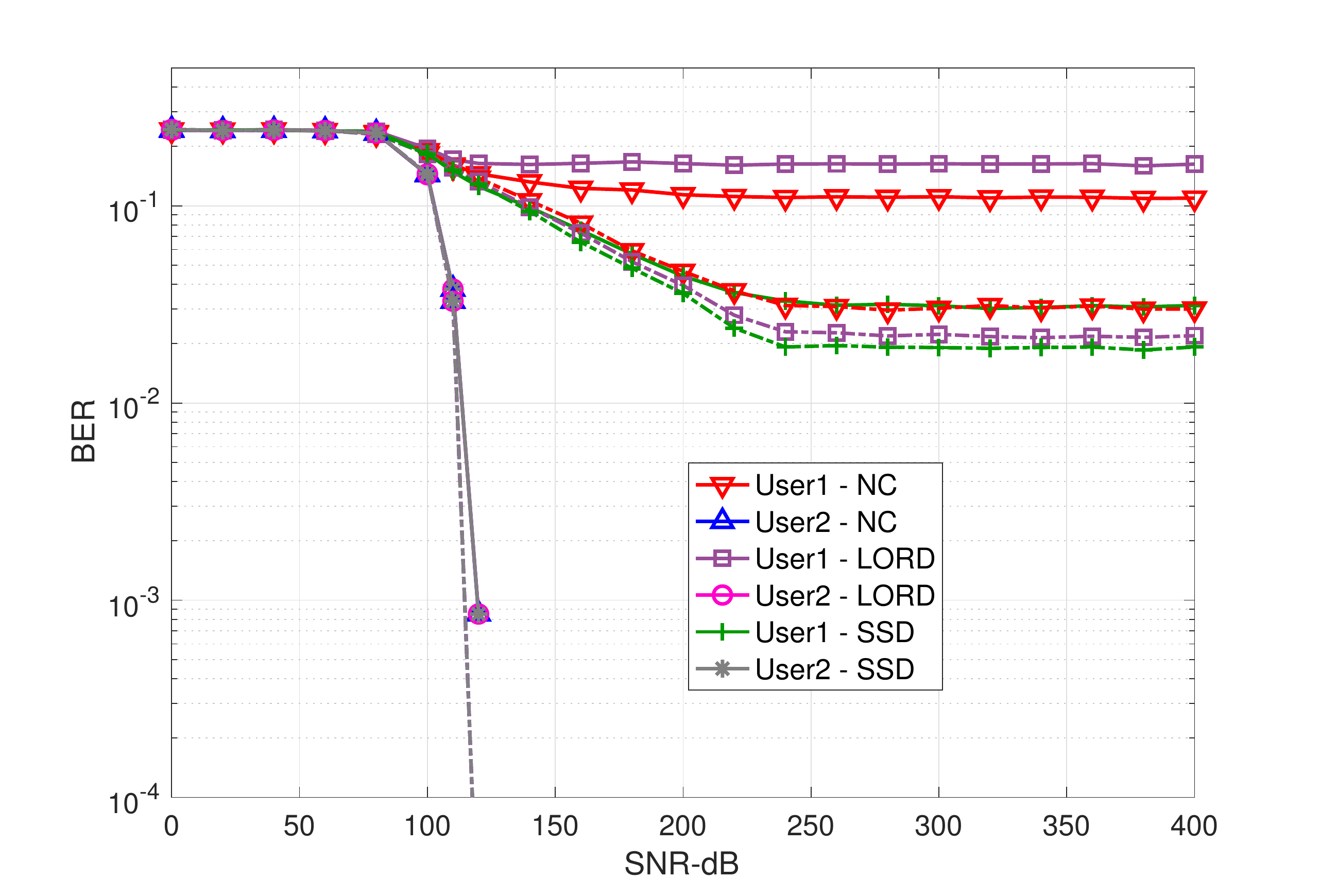}}
  \hfill
  \subfloat[$N=M_{1}=M_{2}=16$ - THz channel with spatial tuning for near user.]{\label{ber1} \includegraphics[width=0.48\linewidth]{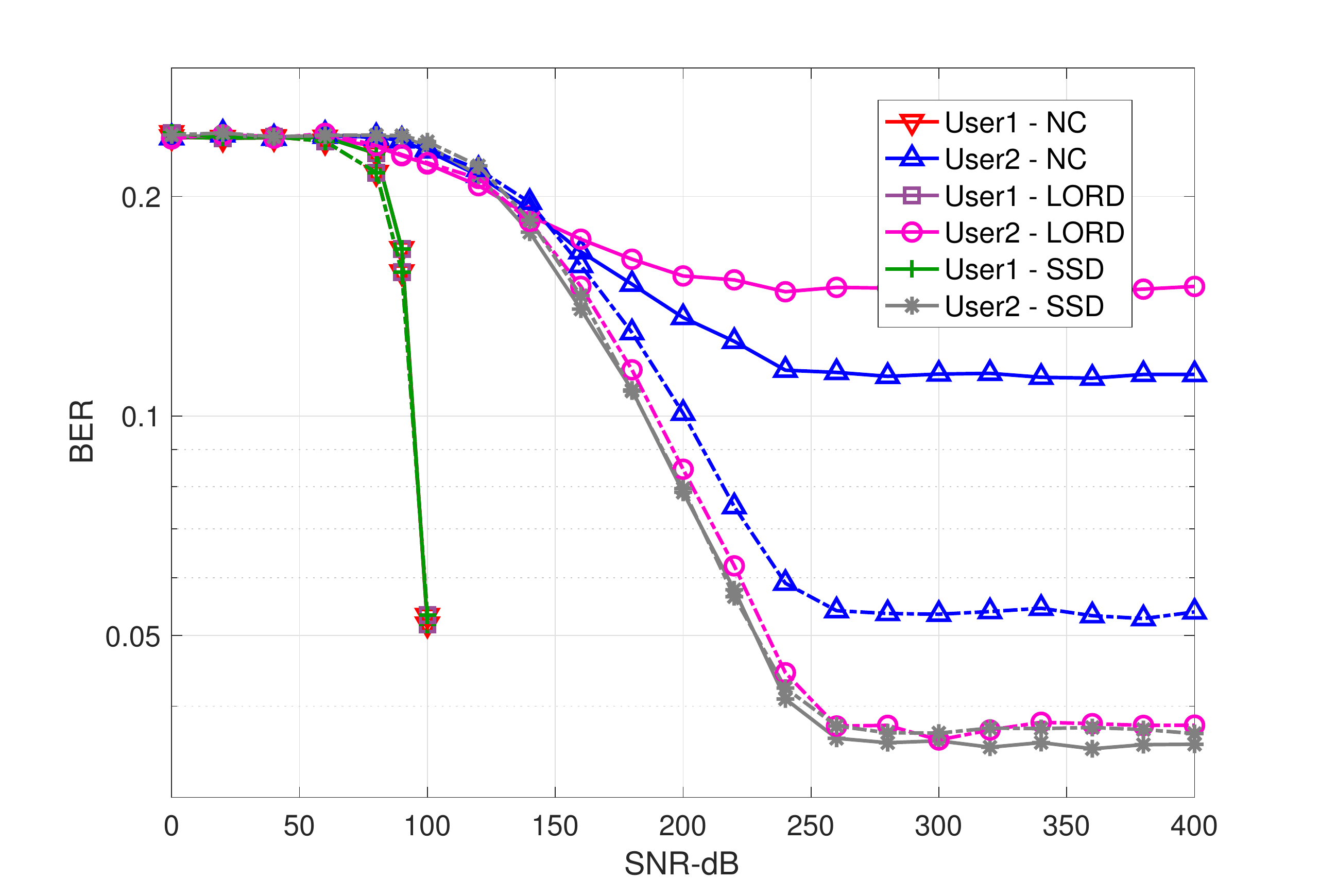}}

  \caption{BER performance of proposed detectors in the multi-user scenario with 16-QAM (dotted lines refer to the reference optimal power allocation scheme).}
  \label{sim_results_2}
\end{figure*}

The complexity tradeoffs of the proposed detectors are particularly important for THz systems. Although THz communications promise Tbps data rates, state-of-the-art baseband clock speeds are confined to a few GHz \cite{sarieddeen2020overview} (1000 bits need to be processed per clock cycle). Furthermore, THz baseband processing capabilities are limited, and there are no energy-efficient transceivers capable of supporting 1 Tbps. While the main complexity burden comes from channel coding and channel code decoding, data detection can significantly reduce complexity, especially in UM-MIMO scenarios. High parallelism is thus an architectural requirement. The parallelizability of the proposed subspace detectors can be exploited to reduce the frame length at the decoders' input. By splitting the code into sub-blocks corresponding to multiple channel decomposition outputs, each sub-block can be processed on a separate decoding core, reducing complexity and memory usage. However, this comes at the expense of additional calculations to mitigate the loss in performance at the sub-block borders. 


Several extensions can further enhance the performance of the proposed detectors. For instance, we can easily modify the construction to account for generating log-likelihood ratios (LLRs) as reliability information in a soft-output (SO) setting. To generate LLRs in SSD, we decouple the $N$ streams in $N$ steps (assuming a single-user scenario with $S_i\!=\!N$). In each step $t \in \{1,\cdots,N\}$ we calculate the LLRs of the bits of symbol $x_n$ ($n=t$). By exchanging LLRs between detection and decoding blocks, iterative detection and decoding schemes are realized following the ``Turbo principle''.  In particular, the decoder can be fed a priori information $\mathrm{LLR}^{A}$, the difference between the detector's SO and its own SO from the previous decoding iteration. The decoder then generates extrinsic LLRs in the form of a posteriori information denoted $\ \mathrm{LLR}^{E}$, where
\begin{equation}
\mathrm{LLR}_{i, m}^{\mathrm{E}}=\mathrm{LLR}_{i, m}-\mathrm{LLR}_{i, m}^{\mathrm{A}}.
\end{equation}
Although iterative schemes are more complex and naturally ill-suited for Tbps constraints, we can adapt the number of iterations according to the THz channel conditions. We can maintain a trade-off between complexity and performance by favoring detection iterations and lowering the number of decoder iterations, for example. In particular, with inherent parallelizability in our proposed detectors, decoding iterations can be saved from specific sub-decoders and distributed to other blocks for better efficiency. Such adaptive iterative detection and decoding can be complemented by an adaptive transmission scheme (mapping bits to symbols). Note that the subspace detectors themselves can be made iterative \cite{Sarieddeen7418324}.

%
\section{Simulation Results and Discussions}
\label{sec:simulations}

The proposed detectors are simulated according to the system model in Sec.~\ref{sec:sysmodel}. Single-user SC and multi-user MIMO-NOMA scenarios are considered. Fig. \ref{sim_results_1} shows the BER plots for a single-user setting. For reference, all proposed detectors are simulated alongside the ML detector in Fig. \subref*{fig1}, for $4\!\times\!4$ MIMO with QPSK, and assuming two power-multiplexed data streams of the same length ($S_{2}\!=\!S_{1}\!=\!4$) ($p_2/p_1\!=\!1000$). First, we note that the detectors maintain their diversity gains under SC, which means that the power separation is sufficient to cancel residual SIC errors. However, this comes at the expense of a larger SNR span. The best performing detector is SSD, which achieves near-ML performance, followed by LORD, CD, and PCD, respectively (NC and PNC have a diversity order of 1). The BER analysis in Sec.\ref{sec:probBER} validates these results, where PCD is argued to lag behind CD due to performance loss caused by puncturing. Nevertheless, puncturing is argued to result in performance enhancement in SSD compared to LORD.


The results of a THz UM-MIMO scenario where three data streams are multiplexed in a $16\times16$ configuration ($S_{3}\!=\!16$, $S_{2}\!=\!8$, $S_1\!=\!4$) with 16QAM are then shown in figures \subref*{fig2} to \subref*{fig4}. Note that a $16\times16$ configuration at the level of SAs can still be considered an UM-MIMO setting because a very large number of AEs is required in each SA to achieve the required power gains. In \subref*{fig2}, the power separation is in the order of 100 ($p_3/p_2\!=p_2/p_1\!=\!\!100$) and THz spatial tuning is applied. No error floors are noted, which indicates that the power separation successfully decouples the NOMA streams. The best-performing is user 3, and the worst-performing user 1. As expected, all detectors show identical performance under orthogonal channels. In Figures \subref*{fig3} and \subref*{fig4}, for THz multipath and LoS channels, spatial tuning is relaxed, which introduces significant error floors due to channel correlation, despite power separation in the order of 1000 to remove the residual SIC error's impact. The gaps between different detectors are clearer at the user with the highest allocated power. The performance of LORD significantly deteriorates at higher SNR values under severe channel correlation, whereas SSD shows the highest resilience (more than an order of magnitude difference in BER). Note that the observed very high SNR values could be significantly reduced by adding antenna and beamforming gains in UM-MIMO, as argued in Sec. \ref{sec:sp_tuning}. For instance, 1000 AEs per SAs on both transmitting and receiving sides would result in a $\unit[60]{dB}$ SNR gain. 

\begin{table}[t]
\centering
\caption{Simulation Parameters}
\footnotesize
\label{table:1}
\begin{tabular}{l|c}
 \hline
 \textbf{Parameter} & Value   \\\hline
    Cell radius ($R_{\mathrm{C}}$) & $\unit[10]{m}$   \\\hline
    Radius of cell-center disk $C_1$ ($R_{\mathrm{N}}$)  &  $\unit[5]{m}$ \\\hline
    NOMA pairs density ($\bar{\lambda}_1,\bar{\lambda}_2$) & 0.1,\ 0.1 \\\hline
    Average $\#$ NOMA pairs ($K$) & $\mathsf{E}\left[K \right] = \pi\bar{\lambda}_1 R_{\mathrm{N}}^2  $ \\\hline
    Path-loss exponent ($\dot{\alpha}$) & 2.2 \\\hline
    Max. transmit power & $P_{\max} =  \unit[100]{mW} \times N$ \\\hline
	NOMA PC parameter $\mu$  & 10\\\hline
    Receiver sensitivity $\rho_\mathrm{rx}$ & $\unit[-100]{dBm}$ \\\hline
        Frequency & $\unit[300]{GHz}$\\\hline
   \end{tabular}
\end{table}

The BER plots for the multi-user NOMA setting of Sec.~\ref{sec:usecase2} are shown in Fig. \ref{sim_results_2} ($16\times16$ MIMO and 16-QAM), where two users are accommodated per cluster. The simulated THz-specific NOMA system parameters are summarized in Table \ref{table:1}. Three different detectors are tested: NC, LORD, and SSD. The detectors are applied directly at user 2, and successively to detect both symbol vectors at user 1. Four different scenarios are simulated, all of which assume equal antenna numbers at the BS and the two users ($N\!=\!M_{1}\!=\!M_{2}$). The proposed JDPC scheme (solid curves) is compared to a reference optimal power control scheme (dotted curves) \cite{Sun7095538}, which formulates the power allocation problem as an ergodic capacity maximization problem. The optimal PC scheme suffers from high complexity and slow convergence since it employs a bisection search method. On the contrary, our joint clustering and power control scheme has low complexity, and it only relies on the distance-based path-loss parameter for channel inversion. The distance-based path-loss is a very relevant metric since THz channel conditions are highly distance-dependent, and the low-complexity implementation of JDPC is crucial under Tbps baseband constraints. Unlike the optimal approach, our proposed JDPC scheme guarantees more power to the far user in a cluster, which is suitable for SIC. Furthermore, our proposed scheme results in lower power consumption on average, as we have $p_1\!+\!p_2 \!\leq\! \frac{P_{\max}}{N}$, whereas in the optimal scheme the transmission power is always  ${P_{\max}}$, where $p_1\!+\!p_2 \!=\! \frac{P_{\max}}{N}$.

The results for a system where spatial tuning is configured on both users are shown in Fig. \subref*{ber4}. Both optimal power control and channel-inversion-based power control achieve similar BER performances. Furthermore, SSD is clearly shown to outperform LORD at a lower complexity. The superiority of user 1 is also noted. However, tuning SA separations at the transmitter to achieve orthogonality on both channels of both users (at different distances) is not realistic, although optimization schemes can approach such solutions. Figures \subref*{ber3} to \subref*{ber1} illustrate the corresponding results when such tuning is relaxed on either or both of the channels. It is noted that spatial tuning of SA separations is superior to simple power allocation optimization, where the user with an orthogonalized effective channel avoids error floors. In the presence of error floors, SSD schemes are more resilient.  

Finally, it is worth noting that although the achievable gains of power-domain MIMO-NOMA systems are not entirely clear, high-frequency scenarios offer a compelling case for their utilization. In \cite{clerckx2021noma}, the authors argue that MIMO-NOMA solutions can misuse the spatial dimension because they incur a multiplexing gain loss due to fully decoded streams in SIC. In particular, such loss is noted when comparing MIMO-NOMA to other candidate MIMO schemes such as conventional multi-user linear precoding (MU-LP) and newly-proposed rate splitting (RS) techniques, but not when compared to OMA. On the one hand, our proposed efficient SIC subspace detectors can combat this reduction in multiplexing gain. On the other hand, with near-singular THz channels, spatial precoding in MU-LP fails to reduce inter-stream interference. Therefore, the power domain remains a crucial enabler for multiplexing data. 
Moreover, as an extension to this work, and given the importance of IRSs alongside UM-MIMO in THz systems, IRS-assisted multi-beam NOMA techniques can be considered \cite{Wei8422365}; passive IRSs can improve the performance of weak users without requiring additional transmit power.
\section{Conclusions}
\label{sec:conclusions}

In this paper, we propose low complexity subspace detectors for THz MIMO-NOMA systems. We leverage adaptive spatial tuning techniques to allocate NOMA resources and enhance channel conditions. The proposed detectors are studied analytically by deriving approximate error probability expressions and empirically via simulations of single-user (SC) and multi-user scenarios. We propose a low complexity joint clustering and power control scheme that exploits the THz distance-based path-loss parameter to guarantee efficient SIC demodulation. We further present a simple architectural implementation design in which lower-complexity detectors are used as building components of more complex detectors. We demonstrate that the proposed detectors achieve significant parallelism and computational savings at low performance costs, which is much needed for realizing a Tbps baseband for THz communication applications.


\ifCLASSOPTIONcaptionsoff
  \newpage
\fi



\begin{thebibliography}{10}
\providecommand{\url}[1]{#1}
\csname url@samestyle\endcsname
\providecommand{\newblock}{\relax}
\providecommand{\bibinfo}[2]{#2}
\providecommand{\BIBentrySTDinterwordspacing}{\spaceskip=0pt\relax}
\providecommand{\BIBentryALTinterwordstretchfactor}{4}
\providecommand{\BIBentryALTinterwordspacing}{\spaceskip=\fontdimen2\font plus
\BIBentryALTinterwordstretchfactor\fontdimen3\font minus
  \fontdimen4\font\relax}
\providecommand{\BIBforeignlanguage}[2]{{%
\expandafter\ifx\csname l@#1\endcsname\relax
\typeout{** WARNING: IEEEtran.bst: No hyphenation pattern has been}%
\typeout{** loaded for the language `#1'. Using the pattern for}%
\typeout{** the default language instead.}%
\else
\language=\csname l@#1\endcsname
\fi
#2}}
\providecommand{\BIBdecl}{\relax}
\BIBdecl

\bibitem{Akyildiz6882305}
I.~F. Akyildiz, J.~M. Jornet, and C.~Han, ``Tera{N}ets: {U}ltra-broadband
  communication networks in the terahertz band,'' \emph{{IEEE} Wireless
  Commun.}, vol.~21, no.~4, pp. 130--135, Aug. 2014.

\bibitem{sarieddeen2019generation}
H.~Sarieddeen, N.~Saeed, T.~Y. Al-Naffouri, and M.-S. Alouini, ``Next
  generation terahertz communications: {A} rendezvous of sensing, imaging, and
  localization,'' \emph{IEEE Commun. Mag.}, May 2020.

\bibitem{Rangan6732923}
S.~Rangan, T.~S. Rappaport, and E.~Erkip, ``Millimeter-wave cellular wireless
  networks: {P}otentials and challenges,'' \emph{Proceedings of the IEEE}, vol.
  102, no.~3, pp. 366--385, Mar. 2014.

\bibitem{akyildiz2014terahertz}
I.~F. Akyildiz, J.~M. Jornet, and C.~Han, ``Terahertz band: {N}ext frontier for
  wireless communications,'' \emph{Physical Communication}, vol.~12, pp.
  16--32, Sep. 2014.

\bibitem{rajatheva2020scoring}
N.~Rajatheva, I.~Atzeni, S.~Bicais, E.~Bjornson, A.~Bourdoux, S.~Buzzi,
  C.~D'Andrea, J.-B. Dore, S.~Erkucuk, M.~Fuentes \emph{et~al.}, ``Scoring the
  terabit/s goal: {B}roadband connectivity in {6G},'' \emph{arXiv preprint
  arXiv:2008.07220}, 2020.

\bibitem{sengupta2018terahertz}
K.~Sengupta, T.~Nagatsuma, and D.~M. Mittleman, ``Terahertz integrated
  electronic and hybrid electronic-photonic systems,'' \emph{Nature
  Electronics}, vol.~1, no.~12, p. 622, 2018.

\bibitem{6708549Jornet}
J.~M. {Jornet} and I.~F. {Akyildiz}, ``Graphene-based plasmonic nano-antenna
  for terahertz band communication in nanonetworks,'' \emph{{IEEE} J. Sel.
  Areas Commun.}, vol.~31, no.~12, pp. 685--694, Dec. 2013.

\bibitem{akyildiz2016realizing}
I.~F. Akyildiz and J.~M. Jornet, ``Realizing ultra-massive {MIMO}
  ($1024\!\times\!1024$) communication in the (0.06-10) terahertz band,''
  \emph{Nano Communication Networks}, vol.~8, pp. 46--54, 2016.

\bibitem{faisal2019ultra}
A.~{Faisal}, H.~{Sarieddeen}, H.~{Dahrouj}, T.~Y. {Al-Naffouri}, and M.~S.
  {Alouini}, ``Ultramassive {MIMO} systems at terahertz bands: {P}rospects and
  challenges,'' \emph{{IEEE} Veh. Technol. Mag.}, vol.~15, no.~4, pp. 33--42,
  2020.

\bibitem{sarieddeen2020overview}
H.~Sarieddeen, M.-S. Alouini, and T.~Y. Al-Naffouri, ``An overview of signal
  processing techniques for terahertz communications,'' \emph{arXiv preprint
  arXiv:2005.13176}, 2020.

\bibitem{Han7321055}
C.~Han, A.~O. Bicen, and I.~F. Akyildiz, ``Multi-wideband waveform design for
  distance-adaptive wireless communications in the terahertz band,''
  \emph{{IEEE} Trans. Signal Process.}, vol.~64, no.~4, pp. 910--922, Feb.
  2016.

\bibitem{Weithoffer8109974}
S.~{Weithoffer}, M.~{Herrmann}, C.~{Kestel}, and N.~{Wehn}, ``Advanced wireless
  digital baseband signal processing beyond {100 Gbit/s},'' in \emph{2017 IEEE
  International Workshop on Signal Processing Systems (SiPS)}, 2017, pp. 1--6.

\bibitem{Ding7973146}
Z.~Ding, X.~Lei, G.~K. Karagiannidis, R.~Schober, J.~Yuan, and V.~K. Bhargava,
  ``A survey on non-orthogonal multiple access for {5G} networks: Research
  challenges and future trends,'' \emph{{IEEE} J. Sel. Areas Commun.}, vol.~35,
  no.~10, pp. 2181--2195, Oct, 2017.

\bibitem{Dai8357810}
L.~Dai, B.~Wang, Z.~Ding, Z.~Wang, S.~Chen, and L.~Hanzo, ``A survey of
  non-orthogonal multiple access for {5G},'' \emph{{IEEE} Commun. Surveys
  Tuts.}, vol.~20, no.~3, pp. 2294--2323, thirdquarter 2018.

\bibitem{Zhu8798636}
L.~{Zhu}, Z.~{Xiao}, X.~{Xia}, and D.~{Oliver Wu}, ``Millimeter-wave
  communications with non-orthogonal multiple access for {B5G/6G},'' \emph{IEEE
  Access}, vol.~7, pp. 116\,123--116\,132, 2019.

\bibitem{Ulgen9298080}
O.~{{\"U}lgen}, S.~{Erk{\"u}c{\"u}k}, and T.~{Baykas}, ``Non-orthogonal
  multiple access for terahertz communication networks,'' in \emph{2020 11th
  IEEE Annual Ubiquitous Computing, Electronics Mobile Communication Conference
  (UEMCON)}, 2020, pp. 0737--0742.

\bibitem{Zhang8824971}
X.~{Zhang}, C.~{Han}, and X.~{Wang}, ``Joint beamforming-power-bandwidth
  allocation in terahertz {NOMA} networks,'' in \emph{2019 16th Annual IEEE
  International Conference on Sensing, Communication, and Networking (SECON)},
  Jun. 2019, pp. 1--9.

\bibitem{zhang2020energy}
H.~Zhang, H.~Zhang, K.~Long, J.~Dong, V.~Leung \emph{et~al.}, ``Energy
  efficient user clustering, hybrid precoding and power optimization in
  terahertz {MIMO-NOMA} systems,'' \emph{arXiv preprint arXiv:2005.01053},
  2020.

\bibitem{Lin7786122}
C.~Lin and G.~Y.~L. Li, ``Terahertz communications: An array-of-subarrays
  solution,'' \emph{{IEEE} Commun. Mag.}, vol.~54, no.~12, pp. 124--131, Dec.
  2016.

\bibitem{Sarieddeen8765243}
H.~{Sarieddeen}, M.~{Alouini}, and T.~Y. {Al-Naffouri}, ``Terahertz-band
  ultra-massive spatial modulation {MIMO},'' \emph{{IEEE} J. Sel. Areas
  Commun.}, vol.~37, no.~9, pp. 2040--2052, Sep. 2019.

\bibitem{Zhu7982784}
J.~Zhu, J.~Wang, Y.~Huang, S.~He, X.~You, and L.~Yang, ``On optimal power
  allocation for downlink non-orthogonal multiple access systems,''
  \emph{{IEEE} J. Sel. Areas Commun.}, vol.~35, no.~12, pp. 2744--2757, Dec.
  2017.

\bibitem{Ngo_2013}
H.~Q. Ngo, E.~G. Larsson, and T.~L. Marzetta, ``Energy and spectral efficiency
  of very large multiuser {MIMO} systems,'' \emph{{IEEE} Trans. Commun.},
  vol.~61, no.~4, pp. 1436--1449, Apr. 2013.

\bibitem{Yang7244171}
S.~{Yang} and L.~{Hanzo}, ``Fifty years of {MIMO} detection: {T}he road to
  large-scale {MIMOs},'' \emph{{IEEE} Commun. Surveys Tuts.}, vol.~17, no.~4,
  pp. 1941--1988, 2015.

\bibitem{choi2006nulling}
J.~Choi, ``Nulling and cancellation detector for {MIMO} and its application to
  multistage receiver for coded signals: performance and optimization,''
  \emph{{IEEE} Trans. Wireless Commun.}, vol.~5, no.~5, pp. 1207--1216, 2006.

\bibitem{waters2008chase}
D.~W. Waters and J.~R. Barry, ``The {Chase} family of detection algorithms for
  multiple-input multiple-output channels,'' \emph{{IEEE} Trans. Signal
  Process.}, vol.~56, no.~2, pp. 739--747, 2008.

\bibitem{Siti-1}
M.~Siti and M.~Fitz, ``A novel soft-output layered orthogonal lattice detector
  for multiple antenna communications,'' in \emph{Proc. IEEE Int. Conf. Commun.
  (ICC)}, vol.~4, 2006, pp. 1686--1691.

\bibitem{8186206Sarieddeen}
H.~Sarieddeen, M.~M. Mansour, and A.~Chehab, ``Large {MIMO} detection schemes
  based on channel puncturing: {P}erformance and complexity analysis,''
  \emph{{IEEE} Trans. Commun.}, no.~99, pp. 1--1, Dec. 2017.

\bibitem{Mansour9298955}
M.~M. {Mansour}, ``Low-complexity soft-output {MIMO} detectors based on optimal
  channel puncturing,'' \emph{{IEEE} Trans. Wireless Commun.}, pp. 1--1, 2020.

\bibitem{Jornet5995306}
J.~M. Jornet and I.~F. Akyildiz, ``Channel modeling and capacity analysis for
  electromagnetic wireless nanonetworks in the terahertz band,'' \emph{{IEEE}
  Trans. Wireless Commun.}, vol.~10, no.~10, pp. 3211--3221, Oct. 2011.

\bibitem{Lin7036065}
C.~{Lin} and G.~Y. {Li}, ``Indoor terahertz communications: {H}ow many antenna
  arrays are needed?'' \emph{{IEEE} Trans. Commun.}, vol.~14, no.~6, pp.
  3097--3107, Jun. 2015.

\bibitem{Torkildson6042312}
E.~Torkildson, U.~Madhow, and M.~Rodwell, ``Indoor millimeter wave {MIMO}:
  {F}easibility and performance,'' \emph{{IEEE} Trans. Wireless Commun.},
  vol.~10, no.~12, pp. 4150--4160, Dec. 2011.

\bibitem{6800118Wang}
P.~{Wang}, Y.~{Li}, X.~{Yuan}, L.~{Song}, and B.~{Vucetic}, ``Tens of gigabits
  wireless communications over {E-Band LoS MIMO} channels with uniform linear
  antenna arrays,'' \emph{{IEEE} Trans. Wireless Commun.}, vol.~13, no.~7, pp.
  3791--3805, Jul. 2014.

\bibitem{ojard2008method}
E.~Ojard and S.~Ariyavisitakul, ``Method and system for approximate maximum
  likelihood ({ML}) detection in a multiple input multiple output ({MIMO})
  receiver,'' Sep.~10 2008, {US} Patent App. 12/207,721.

\bibitem{2014_mansour_SPL_WLD}
M.~M. Mansour, ``A near-{ML} {MIMO} subspace detection algorithm,''
  \emph{{IEEE} Signal Process. Lett.}, vol.~22, no.~4, pp. 408--412, Apr. 2015.

\bibitem{Abbasi9135643}
N.~A. {Abbasi}, A.~{Hariharan}, A.~M. {Nair}, and A.~F. {Molisch}, ``Channel
  measurements and path loss modeling for indoor {THz} communication,'' in
  \emph{2020 14th European Conference on Antennas and Propagation (EuCAP)},
  2020, pp. 1--5.

\bibitem{3gpp032}
\BIBentryALTinterwordspacing
\emph{{U}niversal {G}eographical {A}rea {D}escription ({GAD}) ({Release} 15)},
  3GPP Std. TS 23.032, Jun. 2018. [Online]. Available:
  \url{http://www.3gpp.org}
\BIBentrySTDinterwordspacing

\bibitem{Kim_2008}
J.~Kim and I.~Lee, ``Analysis of symbol error rates for signal space diversity
  in {Rayleigh} fading channels,'' in \emph{Proc. IEEE Int. Conf. Commun.
  (ICC)}, May 2008, pp. 4621--4625.

\bibitem{Sarieddeen7418324}
H.~{Sarieddeen}, M.~M. {Mansour}, L.~M.~A. {Jalloul}, and A.~{Chehab},
  ``Low-complexity {MIMO} detector with {1024-QAM},'' in \emph{Proc. IEEE
  Global Conf. on Signal and Inform. Process. (GlobalSIP)}, Dec. 2015, pp.
  883--887.

\bibitem{Sun7095538}
Q.~Sun, S.~Han, C.~I, and Z.~Pan, ``On the ergodic capacity of {MIMO NOMA}
  systems,'' \emph{{IEEE} Wireless Commun. Lett.}, vol.~4, no.~4, pp. 405--408,
  Aug. 2015.

\bibitem{clerckx2021noma}
B.~Clerckx, Y.~Mao, R.~Schober, E.~Jorswieck, D.~J. Love, J.~Yuan, L.~Hanzo,
  G.~Y. Li, E.~G. Larsson, and G.~Caire, ``Is{ NOMA} efficient in multi-antenna
  networks? {A} critical look at next generation multiple access techniques,''
  \emph{arXiv preprint arXiv:2101.04802}, 2021.

\bibitem{Wei8422365}
Z.~{Wei}, L.~{Zhao}, J.~{Guo}, D.~W.~K. {Ng}, and J.~{Yuan}, ``A multi-beam
  {NOMA} framework for hybrid {mmWave} systems,'' in \emph{Proc. IEEE Int.
  Conf. Commun. (ICC)}, 2018, pp. 1--7.

\end{thebibliography}
\end{document}